\documentclass{article}%
\usepackage{amsmath}
\usepackage{amsfonts}
\usepackage{amssymb}
\usepackage{graphicx}%
\providecommand{\U}[1]{\protect\rule{.1in}{.1in}}

\begin{document}

\title{A Hilbert space setting for interacting higher spin fields and the Higgs issue}
\author{Bert Schroer\\permanent address: Institut f\"{u}r Theoretische Physik\\FU-Berlin, Arnimallee 14, 14195 Berlin, Germany\\present address: CBPF, Rua Dr. Xavier Sigaud 150, \\22290-180 Rio de Janeiro, Brazil}
\date{August 20, 2014}
\maketitle

\begin{abstract}
Wigner's famous 1939 classification of positive energy representations,
combined with the more recent modular localization principle, has led to a
significant conceptual and computational extension of renormalized
perturbation theory to interactions involving fields of higher spin.
Traditionally the clash between pointlike localization and the the Hilbert
space was resolved by passing to a Krein space setting which resulted in the
well-known BRST gauge formulation. Recently it turned out that maintaining a
Hilbert space formulation for interacting higher spin fields requires a
weakening of localization from point- to string-like fields for which the
d=s+1 short distance scaling dimension for integer spins is reduced to d=1 and
and renormalizable couplings in the sense of power-counting exist for any spin.

This new setting leads to a significant conceptual change of the relation of
massless couplings with their massless counterpart. Whereas e.g. the
renormalizable interactions of s=1 massive vectormesons with s%
$<$%
1 matter falls within the standard field-particle setting, their zero mass
limits lead to much less understood phenomena as "infraparticles" and
gluon/quark confinement. It is not surprising that such drastic conceptual
changes in the area of gauge theories also lead to a radical change concerning
the Higgs issue.

\end{abstract}

\section{Improved short distance scaling from string-localization and massive
vectormesons}

It is well-known that in d=1+3 spacetime dimensions only a finite number of
renormalizable couplings between pointlike fields exist. In fact only
interactions among spin s=0 and s=1/2 fields permit a pointlike covariant
renormalization theory \textit{in a Hilbert space setting}. Another
characteristic property of such low spin couplings is that they maintain their
particle interpretation in the massless limit, which manifests itself in the
absence of zero mass infrared divergencies.

This situation changes for $s\geq1$ (referred to as "higher spin"); a
renormalizable pointlike formulation in Hilbert space is not possible, and
infrared divergencies in the massless limit indicate a breakdown of the
standard field-particle relation. The traditional Lagrangian quantization
parallelism to classical field theories leads inevitably to quantum fields in
an indefinite metric space (more precisely a Krein space), as it is well-known
from the Gupta-Bleuler formalism of quantum electrodynamics (QED) and its more
sophisticated successor, the BRST setting \cite{Be-Sto}.

The advantage of this gauge formalism is that, in spite of its later
refinements, it uses basically a renormalization formalism which has been
known since the discovery of covariant perturbative renormalization theory in
the context of quantum electrodynamics (QED) by Tomonaga, Feynman, Schwinger
and Dyson. It is the best compromise between two opposing properties, the
singular nature of quantum fields on the one hand, and on the other hand the
canonical aspects of the classical Lagrangian field formalism. The Hilbert
space structure of quantum field theory (QFT) played no role in the discovery
of perturbative renormalization; but with the contribution of Gupta and
Bleuler, the awareness about the importance of Hilbert space positivity and
its relation to quantum gauge invariance gradually grew.

The path from the formulation of QED to the modern BRST setting of nonabelian
gauge theory passed through several stages among which the adjustment of
renormalization to Yang-Mills couplings by 't Hooft -Veltman, the functional
Faddeev-Popov reformulation and the Slavnov identities are important landmarks.

Whereas in the classical Maxwell theory the introduction of
\textit{vectorpotentials }associated with field strengths was achieved without
problems and their use in quantum mechanical problems in external
electromagnetic fields became indispensable (the quantum mechanical
Aharonov-Bohm effect), their quantization raised serious conceptual questions,
well-known from the Lagrangian quantization setting of QED. The origin of
these problems is not the particular method of quantization but rather the
clash between pointlike localization of $s\geq1~$massless "potentials" with
the Hilbert space positivity. This is well-known from the $s=1$
vectorpotential $A_{\mu}(x)~$associated with the $F_{\mu\nu}~$field strength;
its s=2 counterpart is the second degree symmetric tensor potential $g_{\mu
\nu}(x)$ associated the the field strength $R_{\mu\nu\kappa\lambda}$ (with the
symmetry properties of a linearized Riemann tensor) and for $s=n$ there are
symmetric tensorpotentials of degree $n~$and associated multi-indexed field
strengths with mixed symmetry properties\footnote{Starting at s=3/$\;\not 2
\;$ there are corresponding zero mass "spinor-potentials" and their associated
"spinor field strengths". In the present paper only integer spins will be
considered.}.

There is another more systematic way of looking at this problem. In the
setting of Wigner's positive energy representation theory of the Poincar\'{e}
group the Hilbert space positivity is taken care of by the unitarity of the
group representations. The Wigner representation space for the massless finite
helicity representations $(m=0,h)$ contains covariant wave functions of the
field strength type; the representation space contains however no covariant
pointlike vectorpotentials $A_{\mu}(x).$ In the $s=1$ massive case both,
pointlike covariant field strengths as well as vectorpotentials $A_{\mu}%
^{P}(x)$ (Proca potential) exist. Using the functorial relation between
covariant wave functions and free quantum fields these findings have
corresponding free field counterparts. In the next section we will return to
this issue.

The lack of a Hilbert space description in a quantum theory is a almost a
"contradictio in adjecto" since the Hilbert space is the foundational property
of any quantum theory be it quantum mechanics (QM) or QFT. \ Without the
Hilbert space positivity there will be no quantum probability and none of the
structural properties of QFT (LSZ scattering theory, the connection between
spin and statistics, TCP, ...) can be derived, nor would it be possible to
formulate a physically meaningful causality and localization property. This is
the only property which has no classical counterpart and hence is not part of
the (Lagrangian) quantization parallelism between classical and quantum field
theory. As already mentioned, the problem starts with zero mass quantum fields
of spin s=1 and becomes worse with higher spin. As the Wigner representation
theory shows, it cannot be blamed on a particular quantization procedure but
rather results from a quantum incompatibility between pointlike localization
of ($m=0,s\geq1$) potentials and the Hilbert space positivity. For massive
pointlike $s\geq1~$fields this incompatibility is hidden behind the more
subtle relation between pointlike nonrenormalizability and breakdown of
pointlike localizability (see later).

The simplest way to see that this problem disappears if instead of massless
pointlike potentials one introduces their covariant stringlike counterpart is
to start from a massless field strength and define a massless covariant
vectorpotential which, instead of being localized on a point, "lives" on a
semi-infinite straight spacelike string
\begin{align}
A_{\mu}(x,e)  &  =\int F_{\mu\nu}(x+\lambda e),~~e\cdot e=(e,e)=-1\label{st}\\
&  localized~on~x+\mathbb{R}_{+}e\nonumber
\end{align}
By definition $A_{\mu}(x,e)$ it is covariant ($e$ transforms as a vector) and
two stringlocal potentials commute if they are mutually spacelike separated.
This vectorpotential lives in the same Hilbert space as the field strength and
both fields are in the same localization class. The choice of straight strings
guaranties the covariance. Such potentials are not Euler-Lagrange fields and
hence cannot be obtained by Lagrangian quantization.

The Hilbert space positivity which requires their introduction is not an issue
in classical physics; apart from the gauge freedom in the $A_{\mu}$-$F_{\mu
\nu}$ relation, the potential is a classical fields as any others. Quantizing
classical potentials requires the use of the more subtle quantum gauge theory
whose primary purpose is to recover the Hilbert space positivity. The
disadvantage of such a description is that in the presence of interaction with
matter the gauge-invariant local observables do not contain the physical
matter fields which one needs to generate physical states from the vacuum.

In the massive case the pointlike (Proca) potential is in an interesting way
related to its stringlocal counterpart. Starting from a Proca potential
$A_{\mu}^{P}(x)$ we define%

\begin{align}
F_{\mu\nu}(x)  &  \equiv\partial_{\mu}A_{\nu}^{P}(x)-\partial_{\nu}A_{\mu}%
^{P}(x),~~A_{\mu}(x,e)\equiv\int_{0}^{\infty}d\lambda F_{\mu\nu}(x+\lambda
e)e^{\nu},~e\cdot A=0~\label{pro}\\
\phi(x,e)  &  \equiv\int_{0}^{\infty}e^{\mu}A_{\mu}^{P}(x+\lambda
e)d\lambda,~\curvearrowright A_{\mu}(x,e)=A_{\mu}^{P}(x)+\partial_{\mu}%
\phi(x,e)\nonumber\\
\partial^{\nu}F_{\mu\nu}  &  =m^{2}A_{\mu}=:j_{\mu}^{Max},~~Q^{Max}:=\int
j_{o}^{Max}d^{3}x=0\nonumber
\end{align}
where the linear relation between the stringlocal potential, the Proca
potential and the scalar stringlocal $\phi$ is an algebraic consequence. This
relation between three fields in the same localization (Borchers)
class\footnote{Fields in the same Borchers class are known to describe the
same physics; they represent different "field-coordinatizations" of the same
QFT \cite{St-Wi} \cite{Haag}..} will be important in later sections, but for
the present purpose the relevant property is the observation that the Proca
field has the short distance scaling dimension $d_{sd}=2$ whereas the two
stringlocal fields $A_{\mu},\phi$ have $d_{sd}=1.$ The stringlocal scalar
field will be referred to as the "intrinsic escort" of $A_{\mu}(x,e);$ the
reason for this terminology will become clear in the next section.

The third line defines the identically conserved free Maxwell current whose
associated charge vanishes. This property continues to hold for interacting
massive vectormesons coupled to matter, independent of whether the matter is
charged (massive spinor or scalar QED) or neutral (a Hermitian field $H$). It
is known as the \textit{Schwinger-Swieca screening of the Maxwell charge}.
Here the $H$ does not only stand for Hermitian but also for Higgs. This
screening property and not the metaphor of a spontaneous mass-creating
symmetry breaking referred to as the "Higgs mechanism" is the true intrinsic
property of all massive vectormesons. Complex (charged) matter has also a
conserved charge-anticharge counting current (which is absent for $H$-matter).
The two currents coalesce in the massless limit (in which $H$-interactions
vanishes), More in section 5\textbf{.}

As the spacetime point $x,$ the string direction $e~$is a spacetime parameter
in which the field fluctuates, in fact it may be viewed as a spacetime
position in the unit $d=1+2~$de Sitter space of spacelike directions. These
fluctuations relieve the pointlike fluctuations of the Proca field; formally
the $d_{sd}=2~$derivative of the intrinsic escort field compensates the
leading short distance singularity of the Proca field at the price of a mild
weakening of localization from point- to stringlike. The stringlocal potential
is the only covariant localized vectorpotential which is consistent with the
Hilbert space positivity and localizability in the presence of interactions
(renormalizability$\simeq$pointlike~localizability), in fact it "lives"
together with $F_{\mu\nu}~$and its mutually local Proca sibling in the same
Hilbert space. It is the only \textit{covariant} potential which survives in
the massless limit; the zero mass radiation gauge potential which lives in the
same Hilbert space is neither covariant nor localized. All these
$m>0~$potentials, including the intrinsic escort $\phi$ are linear
combinations of the three $-1\leq s_{3}\leq1$ Wigner creation/annihilation
operators $a^{\#}(p,s_{3})~$with different $u$-$v~$intertwiners.

It is rather straightforward to generalize the idea of stringlocal $d_{sd}%
=1~$(independent of $s$\footnote{In the case of halfinteger $s$ the
s-independent dimension of the stringlocal field is $d_{sd}=3/2.~$In the
present work $s$ is always integer.}); in this case the spin $s~$%
tensorpotentials (conveniently described in terms of a symmetric tensor of
degree $s$) which generalizes the Proca potential has $d_{sd}=s+1~$is
accompanied by $s$ intrinsic escort fields with spins between zero and
$s-1$~which iteratively peel off the leading short distance behavior of the
pointlike $d_{sd}=s+1~$tensor potential. In this paper we will present the
dynamical use of this idea for $m>0,~s=1$ which results in a renormalizable
interacting stringlocal formulation in Hilbert space which, different from the
BRST gauge theory, contains in addition to the local observables the physical
matter fields and permits to pass to $m\rightarrow0$ where only correlation
functions of \textit{stringlocal} physical matter fields survive.

Although stringlocal fields have the $s$-independent short distance
$d_{sd}=1~$and hence permit the construction of interactions within the
power-counting restriction for any spin, only those interactions for which
there exists a pointlike generated vacuum sector of local observables are of
physical interest. In the new setting stringlocal fields are the basic fields
in terms of which the perturbative interactions are formulated; the pointlike
local observables are pointlocal composites of these fields. The main physical
role these stringlocal physical (= acting in Hilbert space) matter fields play
in models of massive vectormesons is that their application to the vacuum
generates states whose large time behavior in the setting of the LSZ
scattering theory approach multi-particle scattering states. On the level of
particles the difference between point- and string-like disappears; The
$e$-dependence of the fields simplifies in the large time limit where it only
leaves its trace in the particles states from where it can be removed by a
change of normalization\footnote{This is similar to the change of one-particle
normalizations in the Haag-Ruelle scattering which is necessary to extract
$\left\vert p,\cdot\right\rangle $ particle components from states generated
by local observables.}. Note that all this changes in the limit $m\rightarrow
0~$of massless vectormesons where scattering theory and together with the
relation between fields and Wigner particles breaks down.

One of the strongest nonperturbative results which relates particles to
stringlocal fields is a theorem by Buchholz and Fredenhagen which states that
in theories with a mass-gap and a nontrivial vacuum sector (generated by the
local observables) all particles (including those which are outside the vacuum
sector) and their scattering states can be generated by operators which are
localized in arbitrary narrow spacelike cones \cite{Bu-Fr}.

The theorem does not In fact our results show that massive QED are the
simplest models Since LSZ scattering theory the proof of this theorem uses the
Hilbert space positivity in an essential way, none of the theorems of QFT can
be expected to hold in a Krein space. In the local quantum physics (LQP)
setting in which this theorem was proven the arbitrary narrow spacelike cones
correspond to semi-infinite spacelike stringlocal fields (they live on the
core of such cones); pointlocal fields correspond in LQP to the core of
arbitrary small double cones and are a special case of (e-independent)
stringlocal fields.

According to this theorem models of QFT with a mass gap are generated by
stringlocal fields $\Psi(x,e);$ pointlocal fields correspond to $e$%
-independent $\Psi.$ Fields which are nonrenormalizable in the pointlike sense
may be renormalizable after converting their first order pointlike interaction
into a suitably defined stringlike form (section 3). This conversion will be
explicitly presented for in the $s=1$ massive QED; it shows that the pointlike
nonrenormalizabilty is caused by the singular nature of the fields which
cannot be localized by smearing with finite supported testfunctions in the
sense of Wightman (they are not operator-valued Schwartz distributions). A
formal pointlike localization can only be attained in Krein space gauge
setting caused at the price the physical meaning of causal localization (which
is only recovered in the vacuum sector of gauge invariant observables). This
shows the powerful role of Hilbert space positivity in relating the breakdown
of pointlike renormalizability with a weakening of localization. It limits the
use of pointlike localization to renormalizable interactions between $s<1$ fields.

The use of Krein spaces as in the BRST gauge formulation is limited to formal
manipulations in perturbation theory, without Hilbert space positivity
locality is physically void and the known properties of the connection between
fields and particles (e.g. spin\&statistics, LSZ scattering theory,..); The
properties of QFT are limited to the gauge invariant vacuum sector for which
the Hilbert space setting is recovered. Quantum gauge symmetry is not a
physical symmetry but rather a mathematical trick to rescue the Hilbert space
of the vacuum sector which is generated by gauge invariant local observables
from an unphysical Krein space setting. In classical field theory the Hilbert
space positivity is not an issue; vectorpotentials are classical fields and
the gauge transformations define a symmetry which transforms between
vectorpotentials associated with a fixed field strength.

In the massless QED limit the the singular pointlike fields disappear and the
strings become rigid; the appearance of infinitely extended photon clouds
along the e-direction convert particles into infraparticles for which the
$e$-direction is "frozen", namely there are no unitaries which change the
direction $e$ (as it was still possible in the massive case) which causes the
spontaneous breakdown of Lorentz invariance \cite{Froe}. These facts can be
derived from the appropriately formulated quantum Gauss law \cite{Bu}.

Interactions involving stringlocal higher spin ($s\geq1$) fields with first
order interaction densities of short distance dimensions $d_{sd}^{int}\leq4$
are renormalizable in the sense of the power counting criterion, whereas their
pointlike analogs are nonrenormalizable since $d_{sd}^{int}(s)>4~$invreases
with $s$. Here "nonrenormalizability" refers to subsumes two properties, on
the one hand the large momentum behavior is not polynomially bounded but
rather increases with the perturbative order; this means that the
corresponding fields cannot be Wightman fields (operator-valued Schwartz
distributions); if they exist at all they are extremely singular with
unbounded short distance dimension $d_{sd}=\infty.$ What makes
nonrenormalizable theories rather useless (apart from phenomenological
applications) is that the growth of the polynomial degree in momentum space is
accompanied by an increase of free counter-term coupling parameters; a theory
which depends on infinitely many parameters looses its predicitive power and
raises doubts about its mathematical consistency.

A renormalizable stringlocal situation of a theory which is pointlike
nonrenormalizable cannot improve the singular pointlike behavior, but at least
it presents the unbounded increase of parameters; the pointlike field is a
singular coordinatization of the same theory with the same finite set of
coupling parameters as those which parametrize the stringlocal renormalizable
description. It will be shown that the higher order stringlocal interaction
densities allow to construct pointlike counterparts for which a direct
construction fails; the so constructed pointlike densities have the expected
bad high energy behavior but contain no new parameters. Since the pointlike
S-matrix can be shown to be identical to its stringlike definitions, this has
the interesting consequence that an S-matrix may have a better high energy
behavior than one would expect on the basis of the LSZ reduction formula which
relates it to the mass shell restriction of singular pointlike fields.

Although in the present work we will limit our constructions to the S-matrix,
the construction of stringlocal fields suggests that the matter counterpart of
the additive relation between massive point- and stringlocal fields
(\ref{pro}) is a multiplicative formula of the exponential form \footnote{All
operator products are formal; their precise meaning in terms of normal
products is part of the problem of their perturbative construction.}%
\begin{equation}
\psi(x)=e^{-ig\phi(x,e)}\psi(x,e) \label{p1}%
\end{equation}
Here the singular (nonrenormalizable) pointlike (spinor, scalar) matter field
is defined in terms of its stringlike sibling; $\phi$ is the stringlocal
scalar escort field and $g~$is the$.$coupling of the massive vectormeson to
the matter field of (scalar, spinor) massive QED. Such exponential relations
are known to convert Wightman fields into singular fields, in fact the
standard illustration is the exponential of a scalar free field \cite{Jaffe}.
Whereas the fields $\psi,\phi$ on the right hand side are Wightman fields
(i.e. can be smeared with Schwartz testfunctions $f(x)h(e)$), this property is
lost in the normal product; such a nonlinear relations do not allow to convert
testfunction spaces.

The relations (\ref{pro}) together with (\ref{p1}) looks like a gauge
transformation. But the conceptual role is different; whereas gauge
transformations are a tool which one needs in order to extract physical
properties (local observables) from an unphysical Krein space description,
their present role is to relate the renormalizable stringlocal description
with its singular pointlike counterpart. Although the calculations are done in
the renormalizable stringloca field (SLF) setting, the relation to the
pointlike objects is important for the construction of the first order
stringlocal interactions and to secure the $e$-independence of the scattering
amplitudes in higher perturbative orders.

Proposals to define gauge invariant matter fields begun to appear shortly
after the dicovery of field quantization in the conrext of QED. Contrary to
gauge invariant local observables (field strength, conserved currents) the
requirement of gauge invariance imposed on matter fields led inevitably to
stringlocal expressions un terms of gauge-variant fields%
\begin{equation}
\psi(x,e)=\psi^{K}(x)e^{ig\int_{0}^{\infty}A_{\mu}^{K}(x+\lambda e)d\lambda
},~~(e,e)=-1, \label{p2}%
\end{equation}
Here the superscript $K$ refers to the fact that \textit{quantum} gauge theory
is formulated in an indefinite metric Krein space and only gauge invariant
fields can be accommodated in a Hilbert space. There is a formal similarity
between (\ref{p1}) and (\ref{p2}) but for our purpose it is more instructive
to emphasize the conceptual differences. In the gauge theoretic representation
one defines a stringlocal physical (gauge invariant) object in terms of
pointlike gauge-variant fields. Our relation (\ref{p2}) on the other hand
defines a very singular field (well-defined in every order with increasing
$d_{sd}$) in terms of a renormalizable stringlocal field. The singular nature
is the price for staying in a Hilbert space. The reason for stressing this
difference is that the similarity of the two formula hides an enormous
conceptual difference which one is not aware of if one remains in the
perturbative realm. Perturbation theory is mainly combinatorics + Feynman loop
integration, but for functional analytic calculations proofs in QFT one needs
the Hilbert space setting, not to mention the probability interpretation of
QT. Formulas as (\ref{p2}) which involve nonlocal composite are outside
computational control whereas the stringlike fields in (\ref{p2}) are the
basic objects of a new Hilbert space based renormalized perturbation theory;
this new setting also permits to calculate pointlike fields which fulfill
(\ref{p1}) in every order perturbation theory.

The new setting leads to a radical change in the way the relation between
massless and massive gauge theories is viewed. In the past, and to a certain
extend even at present, the massless models are viewed as being simpler than
their massive counterpart. This has historical roots, since QED was the first
QFT and models involving massive vectormesons (viz the Higgs model) appeared
much later. But this alleged simplicity of massless models only refers to
formal perturbative properties. The problem starts when one tries to extract
physical properties (infraparticles, confinement). In that case one runs into
nearly intractable "infrared problems" under the rug of perturbation theory
which are related to fundamental changes in the field-particle relation. There
is no such problem in the presence of mass gaps; in this case all the textbook
knowledge about the field-particle relations applies; the LSZ scattering
theory leads to the standard relations between fields and Wigner particles
provided the operators act in a Hilbert space. Only in a Hilbert space setting
the (necessarily stringlocal) electron field is physical and can explain the
infrared properties in terms of long distance behavior of stringlocal matter
fields. The conceptual simplicity of theories with a mass gap suggests to turn
the historical massless-massive relation from its head to its feet by studying
massless interacting vectormesons in terms of massless $s=1~$of their simpler
massive counterparts.

A convincing illustration of this new point of view is provided by the Higgs
model\footnote{For our critical view it is enough to consider the simplest
(abelian) Higgs model.} which in the new setting is simply the renormalizable
coupling of a massive vectormeson to a scalar Hermitian matter field. There
are three types of couplings of massive vectormesons: couplings to charge
(complex) matter (massive QED), to neutral (Hermitian) matter and
self-couplings (unconfined massive gluons and quarks). Even if this is not
evident from the way in which the Higgs model is constructed in the
literature, it must be one of these couplings because the list is exhaustive.
Its standard construction starts from the two-parametric massless scalar QED
(besides the vectorpotential coupling $g$ there is a renormalization-induced
$\left\vert \varphi\right\vert ^{4}$ self-coupling with an independnet
coupling strength $c$). The symmetry (the gauge symmetry !) is spontaneously
broken by a numerical shift $\varphi\rightarrow\varphi+d$ in field space of
the gauge-variant scalar $\varphi$-field; the result is the famous Mexican hat
potential. The parameters $c$ and $d$ are then converted into more physical
mass parameters of the vector meson and $H$- mass $m,m_{H}~$and the result is
presented as a a result of the Higgs-mechanism namely a "mass creation by
spontaneous symmetry-breaking".

The correct presentation, consistent with the conceptual properties of
QFT\footnote{The principles of QFT are expected to determine the masses of
bound states which are generated by composite fields, but the masses of the
interaction-defining "fundamental" fields are part of the definition of a
model.}, is in terms of a first order trilinear $gA\cdot AH$ coupling. Usually
one does not count the masses of the interaction-defining fields as
parameters; this unusual bookkeeping is only necessary if one wants to compare
the resulting theory with the afore mentioned symmetry breaking description
based on the Mexican hat potential which arises from two-parametric scalar QED
through a field shift. Implementing however the gauge invariance of the
S-matrix up to second order $g^{2}$ in the BRST nilpotentent $s$-formalism
($sS=0$)~for the $A$-$H$ model, one finds that the this implementation
\textit{induces quadrilinear self-couplings} with induced coupling strengths
which are mass ratios of the two involved masses $m_{H}^{2}/m^{2}~$which can
be written in terms of a Mexican hat potential. This shows that behind the
alleged mass generation in terms of a metaphoric (gauge !) symmetry breaking
through a shift in field space of (two-parametric) scalar QED there is
something quite different. Instead of gauge symmetry breaking by an imposed
field shift in scalar QED (the Mexican hat potential) the implementation of
BRST gauge symmetry on a $A$-$H$ coupling of a massive vectorpotential to a
Hermitian field \textit{induces}\footnote{Induced couplings are counterterms
with numerical coefficients which are uniquely determined in terms of the
first order data.} a potential of that form.

There are several episodes in theoretical physics in which (the
conceptual/mathematical) nature asserted itself against human misconceptions,
but the Higgs mechanism of mass creation is one of the strongest illustration
because it endured more than 4 decades. The second order calculation presented
in section 5 confirms this observations: \textit{the Mexican hat potential is
not the result of a gauge symmetry breaking mechanism in scalar QED but is
induced from the implementation of BRST gauge invariance of the S-matrix in a
renormalizable }$A$\textit{-}$H$\textit{ coupling of a massive vectormson to a
Hermitian field}.

It is an interesting question to ask why this much simpler description has
been overlooked for such a long time. The answer may be that this coupling has
no massless long range classical limit; the $m\rightarrow0~$limit decouples
the two fields, so the standard quantization construction starting from a
Maxwell setting was not available. The unfortunate idea that the simpler
massive $s=1~$interactions have to be understood in terms of the allegedly
simple massless models accounts for the rest.

Whereas a Goldstone's spontaneous symmetry breaking is characterized by a
conserved current whose associated \textit{charge diverges at long distances}
as the result of a coupled zero mass Goldstone boson\footnote{A shift in field
space on a symmetric $SO(n)$ invariant quadrilinear selfinteractig n-component
scalar field is a mnemonic device to obtain an illustrative example. but the
intrinsic definition is the diverging charge of a conserved current and not
the shift in field space.
\par
{}} \cite{E-S}, massive vectormesons interacting with matter are characterized
by the Schwinger-Swieca screening of the massive Maxwell current. Both cases
represent very different manifestations of conserved currents, the normal case
of a finite nontrivial charge is in the middle between these opposite
extremes. The claimed special mass-giving role (including its own mass) of a
distinguished Higgs particle (the "God particle") is simply the result of a misunderstanding.

From a philosophical viewpoint the removal of a distinguished particle
reestablishes the "nuclear democracy" between particles which carry the same
superselection charge; in the case at hand particle democracy means that there
is no hierarchy between a Higgs particle, a scalar bound state of the escort
field $\phi$ or a scalar massive "gluonium" bound state. The SLF Hilbert space
setting which confirms these results extends this democracy to quantum fields;
instead of having gauge theories in addition to "standard" QFTs, the Hilbert
space setting places all models under the conceptual roof of the causal
localization principle and decides in what cases pointlike field interactions
have to be replaced by their stringlike counterparts (namely for all
ms$\geq1~$interactions).

The formulation of the new Hilbert space setting requires the standard
field-particle relation which is guarantied in the presence of a mass gap
(absence of zero mass fields). Problems of zero mass vectorpotentials must be
approached by taking massless limits of correlation functions involving
interacting stringlocal massive vectormesons which induce stringlocal matter
fields. The singular pointlike matter fields and self-interating massive
gluons disappear in the massless limit and only their stringlocal counterparts
remain. In this SLF setting confinement correspond to the \textit{vanishing of
massless limits of all correlation functions which besides pointlike
observables contain also stringlocal gluon and quark fields}\footnote{The only
expected surviving $q$-$\bar{q}$ configuration is that in which the string
direction is parallel to the spacelike distance between the endpoints of the
$q$ and $\bar{q}~$strings.}. A zero mass situation with peculiar properties is
the noncompact "stuff" associated to Wigner's infinite spin representation
class. Its properties are very different from matter as we know it and only
fit the properties which are ascribed to dark matter. More on this in the next section.

The paper is organized as follows. The next section presents the construction
of stringlocal fields from Wigner's representation theory of the Poincar\'{e}
group with special emphasis of the noncompact "stuff" associated to the third
Wigner class. Section 3 explains the concept of the intrinsic escort field
$\phi~$in more detail, including its role in the model-defining first order
interaction density, as well as its use in the construction of a
string-independent S-matrix. The fourth section contains the simplest
nontrivial second order illustration of the Hilbert space setting of
stringlocal fields (SLF) which is provided by the model of massive scalar QED.
Section 5 presents the analogous second order calculation of coupling of a
massive vectormeson to Hermitian instead of charged matter and its relation to
the Higgs model. The end of section 5 contains additional remarks on
confinement as well as a contrasting juxtaposition of confinement with the
properties of the inert noncompact stuff (dark matter?) associated to the
third Wigner class.

\section{Wigner representations and stringlocal fields}

In the development of the ideas which led up to the present SLF setting, the
Wigner representation theory of the Poincar\'{e} group played a prominent role
\cite{MSY}. There are 3 classes of positive energy representation\footnote{The
positivity of energy is the stability requirement on relativistic quantum
matter.}, the mass $m>0~$class and two zero mass classes: the finite helicity
class and Wigner's "infinite spin" class. With the exception of the third
class, the connection of irreducible positive energy Wigner representations
with interaction-free quantum fields is well-known and has been explicitly
presented in many articles; the most detailed exposition can be found in the
first volume of Weinberg's book \cite{Weinbook}. Written in terms of
$u,v~$intertwiners the result is%
\begin{equation}
\psi^{A,\dot{B}}(x)=\frac{1}{\left(  2\pi\right)  ^{3/2}}\int(e^{ipx}%
u^{A,\dot{B}}(p)\cdot a^{\ast}(p)+e^{-ipx}v^{A,\dot{B}}(p)\cdot b(p))\frac
{d^{3}p}{2p_{0}} \label{Wein}%
\end{equation}
The intertwiners for $m>0$ are rectangular $(2A+1)(2B+1)><(2s+1)$ matrices
which intertwine between the unitary $(2s+1)$-component unitary Wigner
representation and the covariant spinorial representation, and the $a,b$ refer
to particle and antiparticle creation/annihilation operators. For the m=0
representations the formula is the same except that dot stands for the inner
product in a two-dimensional space (the space of the two helicities
$\pm\left\vert h\right\vert $). Another difference is the range of possible
spinorial indices for a given physical spin $s~$the range of spinorial
(half)integer spinorial representation indices of the homogeneous Lorentz
group indices is restricted by%
\begin{align}
\left\vert A-\dot{B}\right\vert  &  \leq s\leq A+\dot{B},~~m>0\label{cov1}\\
\left\vert A-\dot{B}\right\vert  &  =\left\vert h\right\vert ,~~~m=0
\label{cov2}%
\end{align}
the second formula shows that the the vector representation $A=1/2=B$ does not
occur for m=0~i.e. pointlike covariant vectorpotential are not consistent with
the Hilbert space positivity of quantum theory.

The recollection of these facts is helpful for the presentation of the
stringlocal infinite spin field which can be written in the same way
(\ref{Wein}) except that the intertwiner $u$ is not point- but rather
stringlike and that the Hilbert space to which the inner product refers is an
infinite dimensional representation space of the so-called little group (the
full two-dimensional noncompact Euclidean group $E(2)$ which leaves a
lightlike momentum $p$ invariant. In this case the group theoretical method
for the calculation of intertwiners (as used by Weinberg) fails. The solution
came from the calculation of a string-dependent intertwiner using the concept
of "modular localization"\footnote{This new concept is not only important for
the ongoing research in QFT \cite{hol}, but it also permits to understand old
problems in a new light \cite{modular} .} \cite{BGL}. This intrinsic
(independent of field-coordinatization) way of formulating quantum
localization (which had been used before in the construction of integrable
models of QFT \cite{AOP} \cite{integrable} and led to existence proofs
\cite{Lech}), adds a new aspect to Wigner's representation theory. The
positive energy property implies the existence of a nontrivial dense subspaces
of the Wigner space (a kind of ome-particle Reeh-Schlieder property) which
describe spacelike cone-localized wave functions. Whereas for the $m>0$ and
the $m=0$ finite helicity representations the localization can be tightened to
(dense localization spaces for) arbitrary small double cones (which can be
shown to be generated by testfunction smearing of pointlike covariant
wavefunctions), all compact localized subspaces of the third class are
trivial. As a consequence this representation class corresponds to noncompact
"stuff" (in order to distinguish from matter as we know it) which is generated
by stringlocal covariant fields whose field class does not contain pointlocal
composites i.e. the third class Wigner stuff has no local observables).

Since this noncompact localized stuff cannot be approximated by normal
(compact localizable) matter, there is no way to produce it from collisions of
compact matter, nor can it be detected (localized) in (necessarily compact )
earthly counters\footnote{Based on the standard idea that a click in a counter
localizes a particle in the counter (with rapidly decreasing vacuum
polarization caused tails).}. Apart from its stability and its coupling to
gravitation (both properties are shared by all positive energy
representations) it is completely inert. Its natural arena would be galaxies
and its inertness makes it a candidate for dark matter \cite{dark}. Attempts
to define interactions by coupling these fields lead to intractable infrared
problems; unlike interacting zero mass couplings of normal matter there is no
way of resolving infrared problems as limiting cases of models with a mass gap
(as in the case of interacting zero mass vectorpotentials).

The main role of the construction of stringlocal massive fields for $s\geq1$
is to avoid the use of singular pointlike fields which are the cause of
nonrenormalizability in the pointlike Hilbert space setting. Massive
stringlocal fields have some unusual properties. For example stringlocal
scalar fields can linearly (no composites necessary) interpolate particles of
any integer spin, whereas for pointlike massive fields the spin is related to
the spinorial indices (\ref{cov1}). Their unavoidable appearance as intrinsic
escorts $\phi$ of massive vectormesons in stringlike interactions makes them
competitors of extrinsic Higgs fields $H$\footnote{Extrinsic fields add
degrees of freedom whereas intrinsic escorts don't.}$.~$As in the quantum
mechanical condensed matter description of superconductivity where the change
of long ranged vectorpotentials to their corresponding short ranged
counterparts is achieved without the additional degrees of freedom, the
intrinsic escorts $\phi$ permit the existence of massive vectormesons without
adding additional (Higgs) degrees of freedoms.

There is a radical change for zero mass potentials which only exist in the
form of stringlocal fields with appropriate spinorial indices \cite{Pla}.
Whereas their associated pointlike field strengths obey (\ref{cov2}), the
stringlike localization of zero mass potentials permits the wider relation
(\ref{cov1}) between $\left\vert h\right\vert $ and the spinorial indices.
This shows that the algebraic changes in the zero mass limit are very drastic
indeed; the objects on the right hand side of (\ref{pro})
disappear\footnote{Interestingly not only the $\phi~$but also the interacting
$H$ disappears; the zero mass limit of the abelian Higgs model are $A_{\mu},H$
free fields.} and only the stringlocal potential on the left hand side
survives. The representation of the fields changes and the limit can only be
taken for correlations functions from where a Wightman reconstruction
\cite{St-Wi} permits the construction of a new operator formulation. Another
noteworthy property of stringlocal zero mass vectorpotentials which shows
their superiority over the standard indefinite metric pointlike potentials
even in the absence of interactions is the fact that the use of Stokes theorem
in the derivation of the QFT Aharonov-Bohm effect\footnote{Usually this
terminology is used for situations of QM in external electromagnetic fields.}
in the gauge theoretic theoretic setting leads to a zero result, whereas the
stringlocal potential in Hilbert space gives to the correct effect
\cite{charge}. In terms of localization properties the A-B effect is the
special helicity $h=1$ case of a general violation of Haag duality for
multiply connected spacetime regions.\ 

Introducing a semiinfinite line integral $\phi(x,s)$ over the Proca field
along the same spacelike line, one obtains the important linear relation
between the three free fields in the same localization class which share the
same Wigner creation/annihilation operators $a^{\#}(p,s)$ but have different
intertwining functions.

Their two-point functions (including the mixed ones) are consequences of the
properties of the massive Proca field and the above definitions. They can be
computed via the intertwiners, or directly in terms of the above definitions
and the well known two-point function of the Proca field
\begin{align}
\left\langle A_{\mu}^{P}(x)A_{\mu^{\prime}}^{P}(x^{\prime})\right\rangle  &
=\frac{1}{(2\pi)^{3/2}}\int e^{-ipx}M_{\mu\mu^{\prime}}^{A^{P}}(p)\frac
{d^{3}p}{2p_{0}}\label{1}\\
M_{,\mu\mu^{\prime}}^{A^{P}}(p)  &  =-g_{\mu\mu^{\prime}}+\frac{p_{\mu}%
p_{\mu^{\prime}}}{m^{2}}\nonumber
\end{align}
Whereas the short distance scale dimension of the Proca field is $d_{sd}^{P}=$
$2$ (too big for obtaining interactions within the power-counting bounds of
renormalizability), that of the stringlocal potential $A$ is $d_{sd}^{S}=1$%
\begin{align}
M_{\mu\mu^{\prime}}^{A}(p;e,e^{\prime})  &  =-g_{\mu\mu^{\prime}}-\frac
{p_{\mu}p_{\mu^{\prime}}(e\cdot e^{\prime})}{(p\cdot e-i\varepsilon)(p\cdot
e^{\prime}+i\varepsilon)}+\frac{p_{\mu}e_{\mu^{\prime}}}{(p\cdot
e-i\varepsilon)}+\frac{p_{\mu}e_{\mu^{\prime}}^{\prime}}{(p\cdot e^{\prime
}+i\varepsilon)}\label{2}\\
M^{\phi}(p;e,e^{\prime})  &  =\frac{1}{m^{2}}-\frac{e\cdot e^{\prime}}{(p\cdot
e-i\varepsilon)(p\cdot e^{\prime}+i\varepsilon)}\nonumber
\end{align}
where in the second line is the 2-pointfunction of the St\"{u}ckelberg field
and the $\varepsilon~$notation refers to the definition of distributions with
positive energy spectrum in terms of boundary values of analytic
functions\footnote{This and the following e-dependent two-pointfunctions have
been computed by use of the intertwiner functions of the corresponding fields
\cite{MSY} by Mund \cite{Jens1}.}.

The mixed 2-pointfunctions $M^{A,A^{P}},M^{,A,\phi},M^{A^{P},\phi}$ also
follow directly from the definition (\ref{pro})%
\begin{align}
M_{\mu\mu^{\prime}}^{A,A^{P}}(p;e)  &  =-g_{\mu\mu^{\prime}}+\frac{p_{\mu
}e_{\mu^{\prime}}}{(p\cdot e-i\varepsilon)}\nonumber\\
M_{\mu}^{,A,\phi}(p;e,e^{\prime})  &  =\frac{1}{i}(\frac{e_{\mu}^{\prime}%
}{(p\cdot e^{\prime}+i\varepsilon)}-\frac{p_{\mu}e\cdot e^{\prime}}{(p\cdot
e-i\varepsilon)(p\cdot e^{\prime}+i\varepsilon)})\label{3}\\
M_{\mu}^{A^{P},\phi}  &  =i(\frac{p_{\mu}}{m^{2}}-\frac{e_{\mu}^{\prime}%
}{(p\cdot e^{\prime}+i\varepsilon)})\nonumber
\end{align}

The most convenient and systematic way is to use the representation of the
three fields in terms of the $u$-intertwiners which was introduced in
\cite{MSY} and used for the derivation of (\ref{pro}) in \cite{Jens2}. From
the Wightman two-point functions one obtains the stringlocal propagators
\ which will be used in the construction of the $e$-independent S-matrix.

The definition of $d_{sd}=1$ stringlocal$~$tensor fields $A_{\mu_{1}..\mu_{n}%
}$ in terms of their pointlike $d_{sd}=n+1,$ $s=n$ siblings leads to $s$
intrinsic $\phi$-escorts%

\begin{equation}
A_{\mu_{1}..\mu_{n}}(x,e)=A_{\mu_{1}..\mu_{n}}^{P}(x)+\partial_{\mu_{1}}%
\phi_{\mu_{2}..\mu_{n}}+\partial_{\mu_{1}}\partial_{\mu_{2}}\phi_{\mu_{3}%
..\mu_{n}}+...+\partial_{\mu_{1}}...\partial_{\mu_{n_{n}}}\phi\label{escort}%
\end{equation}
The left hand side represents a stringlocal spin $s=n$ tensor potential
associated to a pointlike tensor potential with the same spin. The
$\phi^{\prime}s$ $s=n-i,~i=1,..,n$ tensorial stringlocal fields of dimension
$d=n-i+1$; all the fields belong to the same localization (Borchers) class, in
fact they are linear combinations of the same Wigner creation/annihilation
operators with different intertwiners i.e. the linear relation can be written
as one between intertwiners. Each $\phi$ "peels off" a unit of dimension so
that at the end one is left with the desired spin $s$ stringlocal $d_{sd}=1$
counterpart of the tensor analog of the Proca field. All intrinsic escorts
appear in the first order stringlocal interaction density and play an
important role in the $e$-independence of the S-matrix. \ 

Such relations may be important in attempts to generalize the idea of gauge
theories in terms of SLF couplings involving massive $s>1~$fields. In this
paper we will stay with $s=1.$

\section{Interactions involving stringlocal field}

In this section the idea of stringlike massive vectormeson fields in the same
localization class as their pointlike Proca counterpart is used in order to
convert nonrenormalizable pointlike interactions into stringlike
renormalizable ones. For concreteness take the model of massive (spinor or
scalar) QED. The pointlike interaction density is (all operator products are Wick-ordered)%

\begin{equation}
L^{P}=gj^{\mu}A_{\mu}^{P},~~j^{\mu}=\bar{\psi}\gamma^{\mu}\psi~~or~~j^{\mu
}=\varphi^{\ast}\overleftrightarrow{\partial^{\mu}}\varphi\label{massive}%
\end{equation}
Since the short distance scaling dimension of the massive (Proca)
vectorpotential is $d_{sd}(A^{P})=2,$ the interaction is above the
power-counting limit 4 since $d_{sd}(L^{P})=5.$ Now we use (\ref{pro}) to
rewrite the pointlike interaction in terms of its stringlike counterpart $L$%
\begin{equation}
L^{P}=L\mathcal{-\partial}^{\mu}V_{\mu},~~V_{\mu}\equiv j_{\mu}\phi\label{r}%
\end{equation}

The stringlike interaction density $L$ involves the $d_{sd}(A)=1~$stringlocal
potential $A_{\mu}(x,e)$ and is therefore renormalizable in the sense of
power-counting. It results from the nonrenormalizable $L^{P}$ by "peeling off"
one unit of scaling dimension (for this reading one should bring the
derivative term to the other side) so that $L$ has instead of $5$ only
$d_{sd}=4$. The rewriting of $d_{sc}=5$\ interaction densities into stringlike
renormalizable densities with $d_{sc}=4$ is our construction principle; it
secures in addition to the validity of the power-counting restriction also the
preservation of the physical content which one intuitively associates with
pointlike interaction. In the BRST gauge setting the pointlike
renormalizability is preserved at the price of a loss of Hilbert space. The
Hilbert space positivity which is behind the physical short distance
properties is recovered for gauge invariant operators whereas physical matter
fields of charged particles are outside the range of gauge theory.

Integrating (\ref{r}) with a test function $g(x)~$and taking the adiabatic
limit $g(x)\rightarrow g,$ the divergence term becomes a surface term at
infinity which vanishes in massive models (in the sense of bilinear form
between localized states). Formally the resulting integral represents the
first order S-matrix; since the pointlike and the stringlike expressions
coalesce in the adiabatic limit, this first order S is $e$-independent.

The idea is now to generalize this peeling process (\ref{r}) and the
subsequent adiabatic disposal of high $d_{sd}$ derivative terms in the
adiabatic limit. Since $e$ is a fluctuating variable as $x,$ the number of
$e$-variables increases together with the number of $x.~$The main difference
is that there is no integration over $e^{\prime}s,~$instead the construction
is done in such a way that the $e$-dependence drops out in the adiabatic limit.

For pointlike fields the connection between fields and particles is given in
terms of LSZ scattering theory which leads to the LSZ reduction formula in
which the scattering amplitudes are expressed in terms of mass-shell
restrictions of time-ordered functions of fields. The derivation uses only the
mass-gap property and the Hilbert space positivity i.e. the LSZ theory cannot
be derived in a Krein space BRST gauge setting. It would be extremely
cumbersome to use this formula for perturbative calculations; for that purpose
one uses the St\"{u}ckelberg-Bogoliubov formula which expresses the n$^{th}$
order S-matrix in terms of the time-ordered n-fold product of the first order
interaction density
\begin{equation}
S(gL)\equiv\sum_{n}\frac{i^{n}}{n!}T_{n}(L,\ldots,L)(g,\ldots,g)=:Te^{i\int
L(x)g(x)},~S_{scat}=\lim_{g(x)\rightarrow g}S(gL) \label{S}%
\end{equation}
Here $g(x)\rightarrow g$ is the adiabatic limit in which the spacetime
dependent coupling in the Bogoliubov operator functional $S(g)$ approaches the
physical coupling constant and the S-matrix become Poincar\'{e} invariant.
There is no direct relation between the LSZ reduction formula and the
perturbative Bogoliubov formula. The derivation of the latter uses the
time-development $U(t,s)$ in the interactions picture. Whereas the propagation
operator at fixed times is formal (ill-defined for interacting fields), the
Bogoliubov S-matrix formula is free of these drawbacks.

In order to formalize the idea of $e$ independence of certain objects we
rewrite the relation (\ref{r}) in terms of a \textit{differential form
calculus in the unit }$d=1+2~$\textit{de Sitter space of spacelike directions}
where $L$ and $V_{\mu}$ are zero forms and $Q_{\mu}=d_{e}V_{\mu}$ and
$u=d_{e}\phi$ are exact one-forms%

\begin{align}
&  d_{e}(L\mathcal{-\partial}^{\mu}V_{\mu})=0~~or~L\mathcal{\ }d_{e}%
L=\partial^{\mu}Q_{\mu},~~Q_{\mu}=d_{e}V_{\mu}\label{point1}\\
&  S^{(1)}=\int L^{P}d^{4}x=\int Ld^{4}x~~or~L^{P}\overset{AE}{\simeq
}L\nonumber
\end{align}
i.e. two operators are $d_{e}$-equivalent of their difference is a zero
one-form, and two interactions are adiabatic equivalent (AE) if their
adiabatic limits of their first order S-matrices agree. the two interactions
are adiabatically equivalent (AE).

The multidimensional aspect of this differential calculus appears in higher
orders of the S-matrix. The differential form of the multidimensional higher
order $e$-independence is:%

\begin{align}
&  d_{e}(TLL^{\prime}-\partial^{\mu}TV_{\mu}L^{\prime})=0,~~d_{e^{\prime}%
}(TLL^{\prime}-\partial^{\mu}TV_{\mu}^{\prime})=0\label{point2}\\
&  dTLL^{\prime}-d_{e}\partial^{\mu}TV_{\mu}L^{\prime}-d_{e^{\prime}}%
\partial^{\prime\mu}TLV_{\mu}^{\prime}=0,~d:=d_{e}+d_{e^{\prime}}
\label{point2'}%
\end{align}
where for simplicity of notation, we illustrate the basic idea for n=2 and use
the notation $L^{\prime}$ for $L\mathcal{(}x^{\prime},e^{\prime}\mathcal{)}%
$.~The second line is the manifest symmetric form of (\ref{point2}). These
relations are extensions of (\ref{point1}) which account for the noncommutance
of derivatives with the time-ordering at point- or string- crossings.

This suggests to go one step further and write (with the same reasoning)%
\begin{align}
&  d_{e^{\prime}}(\partial_{\mu}TV^{\mu}L^{\prime}-\partial_{\mu}\partial
_{\nu}^{\prime}TV^{\mu}V^{\nu\prime})=0=d_{e}(\partial_{\nu}TLV^{\prime\nu
}-\partial_{\mu}\partial_{\nu}^{\prime}TV^{\mu}V^{\nu\prime})\label{point3}\\
&  TL^{P}L^{\prime P}:=(TLL^{\prime}-\partial_{\mu}TV^{\mu}L^{\prime}%
-\partial_{\nu}TLV^{\prime\nu}+\partial_{\mu}\partial_{\nu}^{\prime}TV^{\mu
}V^{\nu\prime}),\text{ }\curvearrowright dTL^{P}L^{\prime P}=0 \label{point3'}%
\end{align}
where the last line is obtained from adding the first line in (\ref{point3})
to the first line in (\ref{point2}) and writing the result as $d(..)=0$ where
$d=d_{e}+d_{e^{\prime}}~$is applied to the content of the bracket in the
second line.

For the proof of string-independence of the S-matrix $dS=0~$in second
order$,~$the validity of (\ref{point2'}) is sufficient. But the relation
(\ref{point3'}) does more; it \textit{defines} a second order pointlike
interaction, whose direct definition would run into the standard problems of
pointlike fields with nonrenormalizable interactions. The important point here
is that, different from the direct pointlike renormalization theory, the
conversion of the well-behaved stringlike to the more singular pointlike
interaction density \textit{does not add new undetermined parameters}. Another
important message is that the high energy behavior of scattering amplitudes of
$s\geq1$ interactions is better than what one expects from the mass shell
restriction of nonrenormalizable time-ordered correlation functions. The
on-shell improvement by "peeling off" high derivative terms which the
adiabatic limit removes remains hidden in momentum space.

It is an interesting question whether this idea to start with a pointlike
nonrenormalizable interaction and rewrite it the in terms of stringlike fields
and the divergence of a $V_{\mu}$ term also works for $s>1,$ The precondition,
namely the existence of a linear relation between a pointlike potential with
$d_{sd}=s+1$ and its stringlike $d_{ds}=1~$counterpart together with
$s~$intrinsic escort $\phi^{\prime}s$ of spin zero up to $s-1~$(\ref{escort})
is certainly fulfilled.

Returning to the $s=1$ case, the computation of the S-matrix starts by
expanding the $T_{0}$ product of fields in the bracket (\ref{point2}) into
Wick-products. Our main interest is the 1-contraction component (the tree
approximation). It consists of sums over terms where each term is a
time-ordered propagator two fields multiplied with the Wick-ordered product of
the remaining uncontracted fields. Although the tree contribution in terms of
the $T_{0}$ ordering is a well-defined expression, the above $e$-independence
relations (\ref{point2}) are not fulfilled. The violation is called an
"anomaly"%
\begin{align}
A_{e}  &  =d_{e}(T_{0}LL^{\prime}-\partial^{\mu}T_{0}V_{\mu}L^{\prime}%
)_{1},~A_{e^{\prime}}=d_{e^{\prime}}(...),~~A_{e^{\prime}}(x,e;x^{\prime
},e^{\prime})=A_{e}(x^{\prime},e^{\prime};e,x)\label{an}\\
-A_{e}  &  =d_{e}(N_{e}+R_{e}+\partial^{\mu}N_{\mu,e}),~-A_{e^{\prime}%
}=d_{e^{\prime}}(...),~~N,R,N_{\mu}\text{ }are~local\nonumber\\
&  T_{0}LL^{\prime}|_{1}\rightarrow TLL^{\prime}|_{1}=T_{0}LL^{\prime}%
|_{1}+N_{e}+R_{e},~~T_{0}V_{\mu}L^{\prime}|_{1}\rightarrow TV_{\mu}L^{\prime
}|_{1}=T_{0}LV_{\mu}^{\prime}|_{1}+N_{\mu}\nonumber
\end{align}
Here local means that they are products of $\delta(x-x^{\prime})~$%
functions$~$multiplied with Wick-products of four (point-or string-local) free
fields; the subscript $1~$referes to the 1-contraction component. The notation
$N,N_{\mu}$ indicates that they are normalization terms which can be encoded
into a change of time-ordering. All regular (non delta) terms cancel since for
those one can take the derivative inside the time-ordered product and use the
relation in the second line of (\ref{point1})

The remaining $R$ is quadrilinear in terms of scalar fields, including the
scalar stringlocal intrinsic escort $\phi$ of the vectormeson. It is not
present in massive QED but it appears in models in which massive vectormesons
are coupled to Hermitian scalar fields where it leads to a delta function
modification of the second order $TLL^{\prime}\rightarrow TLL^{\prime}%
+\delta(x-x^{\prime})L_{2}$. In fact the $L_{2}$ turns out to have the form of
the Mexican hat potential except that it is not part of the defining
interaction but induced from the requirement of string-independence of the
S-matrix. The main point in section 5 is to show that behind the incorrect
symmetry-breaking mechanism of the Higgs model there is the coupling of a
massive vectormeson $A_{\mu}$ to a Hermitian scalar field$~H$. The
implementation of the BRST gauge invariance for the S-matrix (or the
$e$-independence in the Hilbert space formulation) \textit{induces} a $L_{2}%
~$quadrininear normalization term in $H$ which has the form of a Mexican hat potential.

From a conceptual viewpoint the difference could not be bigger: instead of an
imposed gauge-symmetry violating Mexican hat potental the implementation of
BRST gauge invariance (or better the $e$-independence) on the second order
S-matrix for a $A$-$H$ interaction induces quadrilinear Mexican hat like
$H$-selfinteraction. There are only 3 types of renormalizable massive
vectormeson couplings: couplings to complex (charged) matter (massive QED),
their neutral counterpart (the $H$-coupling) and self-couplings (massive Y-M)
which leaves no place for a mysterious mass-creating Higgs-mechanism. This
confirms an important point made in the operator setting of the BRST gauge
formulation more than 20 years ago by a group at the university of Z\"{u}rich
\cite{Scharf} \cite{Aste}; for a more recent account see \cite{Garcia}.

Perturbative calculations of the (on-shell) S-matrix in massive $s\geq1$
stringlike QFT are simpler than those of fields and their (off-shell)
correlation functions; for calculation of interacting fields one has to extend
the St\"{u}ckelberg, Bogoliubov, Epstein-Glaser (SBEG) formalism for the
S-matrix (\ref{S}) to fields \cite{E-G}. The matter fields enter the
interaction density of massive QED only in the form of pointlike free
currents. The stringlocal interaction transfers the string-locality of the
vectorpotentials to that of the higher order matter fields $\varphi(x,e).$

The perturbative calculation of the matter field confirms the exponential
relation (\ref{p1}) between the renormalizable stringlocal matter field and
its singular pointlike partner. The analogy with exponential composites of a
free scalar field suggests the exponentials of Wightman fields belong to a
singular class of fields which do not permit a localization by smearing with
arbitrary compact supported Schwartz testfunctions. Such singular pointlike
fields are expected to belong to a field class studied by Jaffe \cite{Jaffe}
who succeed to formulate previous ideas \cite{Ba-S} about a connection between
non-renormalizability and breakdown of the standard localization property in a
mathematical precise way. Their connection with renormalizable stringlocal
fields in the present work assigns to them an interesting role in perturbation
theory. A more detailed perturbative discussion of these singular pointlike
objects (\ref{pro}) in terms of stringlocal Wightman fields will be given in
separate work.

The conceptual content of (\ref{pro}) and (\ref{p1}) is, despite the
resemblance with gauge transformations, very different from the role of the
latter. These equations formalize the adherence of pointlike fields and their
stringlike siblings to the same localization class; they have nothing to do
with a gauge symmetry in Krein space whose only purpose is to rescue physical
local quantum observables from an unphysical indefinite metric setting.
Whereas the BRST gauge setting is consistent with Lagrangian quantization, the
SLF Hilbert space setting does not support a quantization parallelism to
classical fields theory; stringlike fields are not solutions of Euler-Lagrange
equations and cannot be used in functional integral representations.
Fortunately perturbation theory in the form of "causal perturbation"
\cite{E-G} does not depend on a quantization parallelism to classical field
theory; an interaction density can be defined in terms of any form of free
fields. The umbilical quantization cord of $s\geq1~$QFT with classical fields
is cut because the clash between the Hilbert space positivity with pointlike
localization which leads to SLF has no counterpart in classical field theory.
The present work shows that this has its strongest impact in gauge theory.

SLF is what the foundational causal localization principle leads to if one
does not force it to pass through the quantization parallelism to the less
fundamental Lagrangian quantization. Classical field theory shares many
analogies with QFT; after all this is the reason why Lagrangian quantization,
patched up by renormalization theory and some additional hindsight, turned out
to be useful. But it looses its guiding power when it comes to structures
which are in contradiction with positivity requirements of the Hilbert space
setting of QT and this includes all $s\geq1$ interactions.

The new SLF Hilbert space setting is particularly simple for models with a
mass gap since in such cases the standard scattering relation between fields
and particles holds and the connection with (singular) pointlike fields is not
completely lost. The S-matrix turns out to be a global $e$-independent
invariant of the local equivalence class of fields which includes
renormalizable stringlocal fields and singular pointlike fields. As mentioned
before the new view inverts the conceptual relation between massive and zero
mass interactions of vectormesons with matter fields and puts it from its head
to its feet. Theories with mass gaps are by far the simpler models since they
follow the standard spacetime LSZ asymptotic relation between fields and
particles, whereas most of the not understood properties, as gluon/quark
confinement and a \textit{spacetime description} of the momentum space recipe
of photon-inclusive cross sections for collisions between charged particles in
QED, refer to massless vectormesons. Physical matter fields coupled to zero
mass vectorpotentials are inherently stringlocal. The only known way to
overcome the problems of the radical change of the Wigner-Fock space in the
massless limit is to study the massless limit of correlation functions und use
the Wightman reconstruction to return to an operator description in an
appropriate Hilbert space. This leads to a much clearer idea of what
confinement is about and suggests perturbative resummation techniques to prove
it (end of section 5).

The new formulation for interacting vectormesons leads to many conceptual and
computational changes which cannot be accomodated in Feynman rules; this
includes the process of \textit{induction} of interactions from
string-independence of scattering amplitudes. The biggest surprise arises from
the application of the new ideas to the interaction of a massive vectormeson
with a Hermitian scalar matter field i.e. the neutral counterpart of massive
scalar QED. All interactions of massive vectormesons in a Hilbert space
setting involve intrinsic escort fields $\phi$ which explicitly participate in
the interaction$.~$It looks like a curious accident that these fields have
many properties which are incorrectly ascribed to $H$-fields. They are
inexorably connected with massive vectormesons and disappear in the massless
limit. But they do not add new degrees of freedom to massive vectormesons nor
do they break any symmetry and create masses. Their appearance is a new
phenomenon of massive $s\geq1$ interactions in a Hilbert space formulation
i.e. their existence is a result of the powerful Hilbert space positivity
which has only partially been taken care of in the Krein space gauge setting.

The SLF Hilbert space formulation does not only reinstate the particle
democracy of the old S-matrix setting (no distinguished mass-giving "God
particle"), but it also removes the apartheid between gauge theories and non
gauge theories by collecting all QFT, independent of their spin, under the
conceptual roof of the foundational causal locality principle of QFT.

A gauge theoretic formulation for $s>1$ interactions does not seem to be
known. On the other hand the higher spin analogs of the linear relation
between stringlocal free fields and their pointlike partners in the same
localization class are rather straightforward; instead of a single scalar
stringlike St\"{u}ckelberg field one obtains a linear relation involving a
family of intrinsic tensor St\"{u}ckelberg fields for all spins up to $s-1$
\cite{modular}. These are the prerequisites for an extension to higher spin
interactions. The difficult problem is to find interactions with lead to local
observables. Since in the present paper our main interest are interactions
involving massive vectormesons, we will not comment on $s>1.$

Since the gauge theoretic setting is an established part of particle theory,
it may be instructive to compare the stringlocal Hilbert space setting with
the gauge theoretic BRST formulation in Krein space in somewhat more detail.
It can be presented in a formally similar manner as (\ref{pro}), namely
as\footnote{Such formulas do not appear in the work of the University of
Z\"{u}rich group \cite{Scharf}; they appear for the first time in
\cite{Jens1}.}%
\begin{equation}
A_{\mu}^{K}(x)\simeq A_{\mu}^{P}(x)+\frac{1}{2m}\partial_{\mu}\phi
^{K}(x),~~\curvearrowright\partial^{\mu}A_{\mu}^{K}(x)-m\phi^{K}(x)\simeq0
\label{Krein}%
\end{equation}
where the superscript $K$ refers again to Krein space i.e. in an indefinite
metric space which is obtained from a Hilbert spaces by changing the metric in
terms of Hermitian operator $\eta$ \cite{Scharf}\cite{Garcia}. Here the
reduction of the short distance scale dimension $d=2$ of the Proca potential
is achieved not by changing the physical localization but rather by "brute
force" namely by compensating the renormalizability-preventing scale dimension
by a free scalar field with the two-point function of the opposite sign so
that the resulting $d=1$ $A_{\mu}^{K}~$potential acts also in Krein space. As
a result of the interaction of this pointlike potential with a $s<1$ matter
field, the indefinite metric creeps into all fields and renders them
unphysical. The difficult part of this formalism is to find an "operator gauge
requirement" which filters out a subalgebra generated by local observable
fields which applied to the vacuum create a Hilbert space.

As well known to the many practitioners of the BRST formalism, this cannot be
done directly since the above relations (\ref{Krein}) cannot be formulated as
operator equations; as they stand they only express equivalences. In order to
obtain a manageable operator formalism one must extend the above set of fields
in terms of anti-commuting ghost operators $u^{K},\hat{u}^{K}.$ The result is
the famous BRST setting in which one can formulate a certain (unphysical)
gauge symmetry in terms of a nilpotent $s$-operation whose only purpose is to
describe the content in terms of local observables as gauge-invariants%
\begin{equation}
sA_{\mu}^{K}=\partial_{\mu}u^{K},~s\phi^{K}=u^{K},~su^{K}=0,~s\hat{u}%
^{K}=-(\partial A^{K}+m^{2}\phi^{K}) \label{BRST}%
\end{equation}
The bracket in the last relations vanishes on physical states. Unlike physical
symmetries gauge symmetries by their very nature cannot be broken (neither
explicitly nor spontaneous) since this would wreck their only purpose, namely
filtering out physics from an unphysical description.

In the above presentation the analogy to a Hilbert space formulation has been
highlighted by choosing the same notation for those operators which permit a
formal correspondence. But the analogy falls apart on the conceptual physical
level since the purpose of abstract cohomological BRST formalism based on a
nilpotent $s$-operation in Krein space is totally different from that of the
differential form calculus $d=d_{e_{1}}+..d_{e_{n}}~$in the space of string
directions (d=1+2 de Sitter spacetime) whose purpose is to relate the
stringlike description to its pointlike counterpart in the same localization
class. The SLF setting retains the pointlike fields $A_{\mu}^{P}$ and the
definition (\ref{p1}) of the singular $\varphi^{P}(x)$ in terms of its
stringlike sibling, whereas the pointlike physical objects, apart from local
observables, are lost in the BRST setting.

A curious side result of this formal analogy is the fact that the SLF
differential form calculus in terms of the $Q_{\mu}$ and $u,$ which
strengthens the formal similarity of the $d$ differential calculus with the
more abstract nilpotent $s$ operation, also allows a better formal
$m\rightarrow0$ limit behavior. Such formal considerations are however no
replacement for the explicit construction of the expectation values of
stringlocal massless theory as massless limits of their stringlocal massive
counterparts. Such constructions, which remain outside the gauge formalism,
provide the safest way of constructing stringlocal physical matter fields in QED.

Another point which remains insufficiently understood is to what extend one
needs gauge symmetry in order to obtain e.g. the relation between the
quadratic second order $A_{\mu}$ coupling and the first order coupling. The
answer is; \textit{one does not need it at all}; the Hilbert space SLF setting
determines the relation between these two couplings (next section). This is
particularly important in Y-M QFT; in that case the typical form of the
nonabelian gauge interaction between stringlocal massive gluons of equal mass
results from locality in conjunction with the Hilbert space positivity. The
relation between the first order defining interaction and the induced local
part of the second order interaction density, which is the epitome of
classical gauge theory, is the conceptual implication of quantum locality and
Hilbert space positivity. Local quantum physics can stand on its own feet; it
does not need the quantization crutches of classical fibre bundle mathematics.
The Lie-algebraic form of the selfinteraction between gluons needs no
classical imposition; it is the result of the foundational causal localization
principle \cite{M-S2}.

The main purpose of the following two sections is the explicit illustration of
these new concepts of SLF in Hilbert space and their computational results in
second order perturbation theory. For a more detailed and mathematically
rigorous presentation of the modifications of renormalized perturbation theory
in the presence of stringlike fields; see the forthcoming work \cite{Jens2}
\cite{Jens3}.

\section{S-matrix of massive scalar QED}

According to the traditional view massless scalar QED is a pointlike model
with two coupling parameter\footnote{The electromagnetic coupling and a
parameter related to a counterterm-induced quadrilinear scalar field
self-coupling.}; it is known to be renormalizable in the pointlike BRST Krein
space setting. Unlike its classical counterpart, this quantum gauge
description is severely restricted; the positivity requirements of the Hilbert
space clash with the pointlike localization and quantum gauge theory is the
result of a compromise. the description is limited to local observables which
constitute the gauge invariant part, whereas the formally gauge-variant
charge-carrying operators and physical charge-carrying states remain outside
the pointlike BRST formalism.

Massiven vectormesons are described in terms of Proca potentials $A_{\mu}^{P}$
whose short distance dimension $d_{sd}=2~$is too high for the construction of
interactions below the power-counting limit. The BRST gauge setting attains
renormalizability by replacing the Proca potentials by $d_{sd}=1$
vectorpotentials $A_{\mu}^{K}$ in Krein space; again the price to be paid is
the rather small physical range of such a description which includes the gauge
invariant local observables but contains no computational accessible
construction for physical matter fields and the states which they create from
the vacuum. In a Hilbert space description the existence of a mass gap would
insure the validity of scattering theory, but without the powerful Hilbert
space positivity the physical compass for navigating through the network of
field-particle relations is lost.

It is the main purpose of the present paper to change this situation by
proposing a Hilbert space setting which maintains the $d_{sd}=1~$short
distance property for any integer spin\footnote{A similar construction with
$d_{sd}=3/2~$is possible for fermions.} and as a result permits to extends
renormalization theory in Hilbert space. Here we are primarily interested in a
Hilbert space setting which replaces the BRST gauge formalism. Since the mass
gap property in Hilbert space setting guaranties the validity of scattering
theory and the standard field-particle relations it seems to be reasonable to
construct massless models as massless limits of their massive counterpart. The
important difference to the BRST gauge setting is that all fields are
physical; there is no point is studying such limits for pointlike
gauge-variant fields. This way of studing massless models as limits of their
massive counterparts is contrary to the standard approach which in its most
extreme form claims that masses of vectormesons should be viewed as arising
from a spontaneous symmetry-breaking from massless vectorpotentials.

The alleged simplicity of massless models refers to formal aspects of
renormalized perturbation theory and ignore the unsolved problems of
"infraparticles" in QED \cite{Bu-Ro}, not to mention the unsolved millennium
problem of confinement in QCD.

In this section the model of massive scalar QED will be used as a nontrivial
testing ground for the new SLF Hilbert space formalism. The idea is to use the
short-distance improvement of stringlike potentials in Hilbert space, while
mainting the pointlike nature of observables which in the massive case
includes the string independence of the S-matrix. It will be shown how the
quadratic term in the vectorponetial arises from the locality of the second
order S-matrix. 

The defining first order stringlocal interaction density of massive scalar
QED
\begin{align}
L(x,e) &  =gA_{\mu}(x,e)j^{\mu}(x)=L^{P}+\partial^{\mu}V_{\mu}\label{L}\\
j^{\mu} &  =\varphi^{\ast}\overleftrightarrow{\partial^{\mu}}\varphi,~V_{\mu
}=\phi j_{\mu}\nonumber
\end{align}
is according to (\ref{point1}) $d_{e}$-equivalent to its pointlocal
counterpart $L^{P}$. This secures the $e$-independence of the first order
S-matrix in the AE limit. In these equivalences the stringlocal
St\"{u}ckelberg field $\phi,$ which appears explicitly in $V_{\mu},$ play an
essential role. Whereas the first order relation is a result of the definition
of a "stringlocal" interaction, the second order relation (\ref{point2}) is a
nontrivial restriction on the renormalization.

One defines a reference time-ordering $T_{0}$ of two-pointfunctions of
derivatives of the complex scalar field $\varphi$ by taking the derivatives
outside the two-point function e.g.%
\[
\left\langle T_{0}\partial_{\mu}\varphi^{\ast}(x)\partial_{\nu}^{\prime
}\varphi(x^{\prime})\right\rangle =i\frac{\partial_{\mu}\partial_{\nu}%
^{\prime}}{\left(  2\pi\right)  ^{4}}\int d^{4}pe^{-ipx}\frac{1}{p^{2}%
-m^{2}+i\varepsilon}%
\]
On the other hand the renormalized time ordering in Epstein and Glaser's
renormalization uses normalization terms whose delta function degree is
determined by the short distance scaling degree of the object to be
renormalized. For the degree 4 propagator of the derivative of a free scalar
field the E-G renormalization leads to the modified $T$-product
\begin{equation}
\left\langle T\partial_{\mu}\varphi^{\ast}(x)\partial_{\nu}^{\prime}%
\varphi(x^{\prime})\right\rangle =\left\langle T_{0}\partial_{\mu}%
\varphi^{\ast}(x)\partial_{\nu}^{\prime}\varphi(x^{\prime})\right\rangle
-aig_{\mu\nu}\delta(x-x^{\prime}) \label{c1}%
\end{equation}
where $a$ is a free parameter.

If we were to treat the defining first order interaction $A_{\mu}j^{\mu}$ in
terms of pointlike $A_{\mu}$ field in a Krein space the interaction is
renormalizable in the perturbative inductive Epstein-Glaser setting where it
leads to two counterterms. The first counterterm (\ref{c1}) appears in the
second order tree approximation and amounts to a modification of the
interaction through a second order contact term (all operator products are
meant to be Wick-ordered)%
\begin{equation}
aA_{\mu}(x)A^{\mu}(x)\varphi^{\ast}(x)\varphi(x) \label{c2}%
\end{equation}
with an \textit{independent} coupling parameter $a.$ There is an additional
quadrilinear counterterm with a coupling parameter of the form%
\begin{equation}
b\left(  \varphi^{\ast}(x)\varphi(x)\right)  ^{2} \label{c3}%
\end{equation}
which appears for the first time in $4^{th}~$order (the box graph); these two
counterterm exhaust the possibilities of E-G counterterm structures
(primitively divergent contributions in the Feynman graph setting), which
means that the renormalized theory is 3-parametric, the first order coupling
and $a,b.$

To recuperate local observables acting in a Hilbert space (at the expense of
charge-carrying matter fields which remain unphysical fields in Krein space)
one has to \textit{extend the Krein space formulation by the ghost operators;}
in this way one arrives at the \textit{BRST gauge formulation which fixes the
parameter }$a$\textit{ in (\ref{c2}) to a numerical value~}$a=2$\textit{ }by
the requirement that the second order S-matrix fulfills $sS=0,~$where $s~$is
the nilpotent BRST s-operation (the BRST implementation of quantum gauge
invariance) according to the rules of a formal "gauge symmetry". By itself
this term (\ref{c2}) has no direct physical interpretation apart from its role
in the extraction of local observables from an unphysical description. An
elegant way to deal with this second order $A\cdot A~$dependent term is to
encode it into the change $T_{0}\rightarrow T$ (\ref{c2}) of the time-ordered
product; in that case the second order tree-contribution rewritten in terms of
$T_{0}~$reproduces this term; in fact the encoding into $T~$covers the tree
component of all orders. The standard gauge formalism leaves the quadratic
contribution and combines it (the gauge-invariant extension) with the first
order term but this does not change the fact that only the sum of $T_{0}LL$
and the pointlike contribution to the second order $S$ is $s$-invariant.

Hence the $s$-invariance reduces the original 3-parametric pointlike model to
two a 2-parameter model, the quadratic second order counterterm is "induced"
by operator gauge invariance. In contrast to classical gauge theory, quantum
gauge symmetry is a technical trick which permits to extract the physics from
a Krein space description; in particular it cannot be spontaneously broken.

In the SLF Hilbert space setting the $e$-independence of the S-matrix induces
the correct value of $a$ from the model-defining first order $A\cdot
j~$interaction; it is simply the result of the implementation of locality in
Hilbert space setting. No additional principle as gauge symmetry has to be
invoked in order to fix $a$ to its correct numerical value. The
\textit{induction mechanism}~exists only for higher spins $s\geq1;$ for lower
spins the renormalization theory is the well-known counterterm formalism with
freely varying coupling strengths.

For the case at hand $a$ is calculated as follows. From the results in the
previous section we know that the second order locality requirement for the
S-matrix in the presence of stringlike fields amounts to the vanishing of the
$d_{e}~$operation on the renormalized tree ($1$-contraction) component%
\begin{align}
&  d_{e}(TA\cdot jA^{\prime}\cdot j^{\prime}-\partial^{\mu}T\phi j_{\mu
}A^{\prime}\cdot j^{\prime})_{1-con}=0\label{c4}\\
-A_{e}  &  :=d_{e}(T_{0}A\cdot jA^{\prime}\cdot j^{\prime}-\partial^{\mu}%
T_{0}\phi j_{\mu}A^{\prime}\cdot j^{\prime})_{1-con}=N_{e}+\partial^{\mu
}N_{_{e},\mu}\nonumber
\end{align}
and a similar expression in which the unprimed and primed $x,e$ are
interchanged. Both $N$ are products of a delta function $\delta(x-x^{\prime
})~$with~a Wick polynomial of degree 4. The simplicity of the model allows us
to take a short cut which bypasses the calculation of the $N^{\prime}s$ in the
anomaly. By inspection on sees that the choice $a=1~$in the definition of the
"renormalized" $T$ (\ref{c1}) solves the problem of the anomalies from
$\varphi$-contractions; as a consequence of the identity $d_{e}\partial^{\mu
}\phi=d_{e}A^{\mu}$ there are no contributions from $\phi$-$A_{\nu}$
contractions. This renormalized $T~$product is characterized by the absence of
the propagator anomaly for the derivative of the $\varphi$-field.%
\begin{equation}
\partial^{\mu}\left\langle T\partial_{\mu}\varphi^{\ast}(x)\partial_{\nu
}^{\prime}\varphi(x^{\prime})\right\rangle =-i\partial_{\nu}^{\prime}%
\delta(x-x^{\prime})-ia\partial_{v}\delta(x-x^{\prime})\text{ }%
+reg=reg~~if~a=1
\end{equation}
The $N_{e}$ and $N_{e,\mu}$ can be red off from the difference between the $T$
and $T_{0}$ in (\ref{c4})$.$%
\begin{equation}
N_{e}=\delta\varphi^{\ast}\varphi A\cdot A^{\prime},~N_{e,\mu}=\delta
\varphi^{\ast}\varphi\phi A_{\mu}^{\prime}%
\end{equation}
The $N_{e^{\prime}},~N_{e^{\prime},\mu}~$result from $e\leftrightarrow
e^{\prime}~$and the $N$ and $N_{\mu}~$are the sums of the primed and unprimed
$Ns.$

As expected from gauge theory, $a=1~$leads to a $N~$which is quadratic in the
vectorpotential of the form\footnote{We remind the reader that all operator
products are Wick-products.}%
\begin{equation}
2\varphi^{\ast}A_{\mu}\delta(x-x^{\prime})\varphi^{\prime}A^{\mu\prime}%
\end{equation}
There is a small but nevertheless important difference to the corresponding
gauge theoretic result; the two $A~$have independently fluctuating string
directions. This is quite different from the use of the axial gauge in gauge
theory where the coalescing gauge parameter causes intractable renormalization
problems which led to the abandonment of this gauge. Together with the
contribution from $N_{e^{\prime}}$ with $x,e\longleftrightarrow x^{\prime
},e^{\prime}$ one finds the $e$-$e^{\prime}~$symmetric form%
\begin{align*}
TLL^{\prime}  &  =T_{0}LL^{\prime}+2i\delta(x-x^{\prime})L_{2},~L_{2}%
=2\varphi^{\ast}(x)\varphi(x)A\cdot A^{\prime}\\
S  &  =ig\int(L+\frac{-i}{2}gL_{2})-g^{2}\frac{1}{2}\int\int T_{0}LL^{\prime
}+higher~orders
\end{align*}
The last line is the gauge theoretic way of writing the result up to second
order. But the preferable notation in the SLF setting is to encode the $L_{2}$
term into a modified $T$-product. The reason is that only the sum in the
second formula leads to a $e$-independent second order S-matrix. The
$T$-encoding instead of the $T_{0}$ has the additional advantage that it takes
care of all the higher order tree contributions. In a formal pointlike setting
all counterterm couplings are independent; the \textit{induction} of $a$ which
reduces the number of freely varying counterterm parameters is the result of
$e$-independence of the second order S-matrix.

Another important difference to the pointlike setting is the possibility to
use the relation (\ref{point3'}) of the previous section to define a pointlike
second order interaction density. Such a calculation is more involved than
that for the S-matrix since one also has to calculate the "renormalized"
derivative terms (the $N_{\mu}$ terms). As mentioned before the SLF Hilbert
space setting induces pointlike densities; the standard problem of
nonrenormalizability of having an ever increasing number of counterterms with
new couplings is evaded although the presence of derivative terms increases
the high energy behavior. Since the derivative terms drop out in the adiabatic
S-matrix limit the scattering amplitudes inherent the improved high energy
behavior of the stringlocal interaction density. This may serve as a warning
against inferring the presence of additional particles on the basis of Feynman
graphs for pointlike interactions.

\section{Maxwell-currents, charge-screening and the Higgs issue}

Additional information about the stringlocal setting for massive QED can be
obtained from the extension of the perturbative formalism to the construction
of stringlocal fields and their possibly pointlike composites. They reveal
aspects which in the global S-matrix remain hidden. One such observable is the
identically conserved Maxwell-current $j$ which is defined as the divergence
of the field strength
\begin{align}
&  \partial^{\nu}F_{\mu\nu}=gj_{\mu},~~F_{\mu\nu}=\partial_{\mu}A_{\nu
}-\partial_{\nu}A_{\mu},~~j_{\mu}=\partial^{\nu}F_{\mu\nu}\label{F}\\
&  J_{\mu}=\frac{1}{2}i\varphi^{\ast}D_{\mu}\varphi+h.c.,~Q=\int J_{0}%
(x)d^{3}x,Q^{Max}=\int j_{0}(x)d^{3}x\nonumber
\end{align}
where the integral over the Maxwell-current $j_{\mu}$ defines the Maxwell
charge $Q_{M}.$ In the massless case this charge coalesces with the
particle-antiparticle "counting charge", but it deviates in a physically
significant way in the presence of massive vectormesons. This difference finds
its physical expression in the \textit{charge-screening} theorem which
confirms a conjecture by Schwinger \cite{Schw}; already the free Proca
Potential leads as a result of $j_{\mu}=m^{2}A_{\mu}^{P}$ to a screened charge
$Q_{M}=0$.

\textbf{Theorem} (Swieca 1976 \cite{Sw}\cite{B-F}) In the presence of a mass
gap, the identically conserved current associated with an antisymmetric tensor
$F_{\mu\nu}~$leads to a screened charge $Q_{M}=\int j_{0}d^{3}x=0.$

In order to avoid any confusion, QFT in this paper always refers to a theory
of quantum fields (or localized nets of algebras \cite{Haag}) in
\textit{Hilbert space}; the proof of this and the following structural
theorems depend on the Hilbert space positivity in an essential way; no
structural theorem in QFT can be derived without the Hilbert space positivity.
There exists another structural theorem which has a close historical
connection to the screening theorem and which is equally interesting in the
context of massive gauge theories and the $H$-coupling.

\textbf{Theorem} (Buchholz-Fredenhagen 1982 \cite{Bu-Fr}) In a QFT with local
observables and a mass gap, charge-carrying matter can be localized in
arbitrary narrow spacelike cones (whose cores are semi-infinite spacelike strings).

This theorem states that, under the mentioned restriction, a QFT can be
generated from operators which are members of operator algebras localized in
arbitrary narrow spacelike cones i.e. objects which are localized on spacelike
surfaces are not needed as building blocks of such a QFT. The possibility of
generating a particular model in terms of objects localized in arbitrary small
double cone\footnote{Narrow spacelike cones are the smallest noncompact
causally closed regions whereas the smallest compact such regions.} regions
(whose core is a point) is a special case covered by the theorem. \ 

It is believed that this localization property continues to be valid in
massless models for $s\geq1.$In the case of QED there exists a structural
proof based on the appropriately formulated quantum Gauss law \cite{Bu}. As a
result of infinitely extended photon-clouds in the $e$-direction, the charged
stringlocal matter fields in QED are more rigid than their massive
counterparts. Whereas the string directions$~$of massive strings can be
changed by Lorentz transformations, the Lorentz invariance in QED is
spontaneously broken \cite{Froe} and the string directions define a continuous
set of superselection rules within which the countable set of charge
superselection rules can be unhinged \cite{Bu-Ro}.

These zero mass localization aspects are inexorably linked with the occurrence
of perturbative logarithmic divergencies in scattering amplitudes whose
resummation in the leading logarithmic is an important tool in the study of
the physics behind perturbative infrared divergencies. The resummation before
taking the massless limit is known to leads to the well-known vanishing of
scattering amplitudes for scattering of charged particles with a finite number
of photons in the in/out states \cite{YFS}. The physical aspects of collision
theory in such cases are described in terms of photon-inclusive cross
sections. Another spacetime interpretation of this result is based on the idea
that charged "infraparticles" have time-ordered correlation functions which
instead of the usual mass-shell poles have milder coupling-dependent cut
singularities. The latter are too mild for being able to compensate the large
time spreading of wave packets in the LSZ scattering limits and consequently
lead to a vanishing $t\rightarrow\infty$ limit. On the other hand zero mass
interaction of pointlike fields with $s<1$ maintain the particle structure and
are consistent with the standard time-dependent scattering theory.

It is useful for later purpose to complete this list of structural theorems by
adding a third theorem about spontaneous symmetry breaking which is a precise
version of Goldstone's idea which he exemplified on Lagrangian models.

\textbf{Theorem}: (Ezawa-Swieca 1967 \cite{E-S}\cite{Cargese}) The
large-distance divergence of the charge associated to a conserved current
(i.e. the intrinsic definition of a spontaneous symmetry breaking)
$Q_{G}=\infty$ is the result of the presence of a massless Goldstone boson
which couples to the current.

Emboldened by the successful test for the stringlocal renormalization setting
for the charged scalar fields in the previous section, we now consider the
coupling of a massive vectormeson to a \textit{neutral field} $H,$ where as
before $H$ stands for Hermitian or (as it will become clear later on) Higgs.
We could call such a coupling massive "chargeless QED" if it would not be for
the fact that its massless limit is trivial since the interaction disappears.

As already indicated before, the renormalization theory for a chargeless
(Hermitian) coupling will lead to more induced (even and odd) terms since the
evenness from (particle-antiparticle) charge conservation is now absent. We
start the induction from the most general pointlike interaction of engineering
dimension $d_{en}=3$ between a Proca potential and a scalar Hermitian field
(omitting the coupling strength $g$)%

\begin{equation}
L^{P}=m(gA_{\mu}^{P}A^{P\mu}H+bH^{3}) \label{defining}%
\end{equation}
where the presence of the factor $m~$(the vectormeson mass) maintains the
engineering dimension to be that of an interaction density, namely
$d_{eng}=4.$ Since the short distance scale dimension of the Proca field is
$d=2,$ the operator dimension of the interaction density is $d=5;$ hence the
pointlike model is nonrenormalizable, as expected. A third possible
$d=5~$trilinear term $A_{\mu}^{P}\partial^{\mu}HH$ does not contribute since
as a result of $\partial^{\mu}A_{\mu}^{P}=0$ it turns out to be a total
derivative. We will not add a quadrilinear term $cH^{4}$ but we will later see
that the string-independence of the S-matrix induces such a term (the induced
Mexican hat potential).

The "peeling formula" for $L^{P}$ i.e. its decomposition into a stringlocal
$L$ and an on-shell disposable "surface term" is straightforward and leads to%

\begin{align}
&  L^{P}=L\mathcal{-\partial}^{\mu}V_{\mu},~with~L=m(A_{\mu}A^{\mu}H+A^{\mu
}\phi\overleftrightarrow{\partial}_{\mu}H-\frac{m_{H}^{2}}{2}\phi^{2}%
H+bH^{3})\label{first}\\
&  and~V_{\mu}=m(A_{\mu}\phi H+\frac{1}{2}\phi^{2}\overleftrightarrow
{\partial}_{\mu}H),~Q_{\mu}=d_{e}V_{\mu}=m(A_{\mu}uH+u\phi\overleftrightarrow
{\partial}_{\mu}H)\nonumber
\end{align}
where in returning from $L$ to $L^{P}$ the $m_{H}^{2}$ in $L$ is compensated
by the mass term in the Klein-Gordon equation which results from the
divergence of $V_{\mu}.$ Let us now look at the second order relation
(\ref{point2}) which expresses the independence from $e~$(with a corresponding
relation for $d_{e^{\prime}}$)%
\begin{equation}
d_{e}(T_{0}LL^{\prime}-\partial^{\mu}T_{0}V_{\mu}L^{\prime})_{1-con}%
=(d_{e}T_{0}LL^{\prime}-\partial^{\mu}T_{0}Q_{\mu}L^{\prime})_{1-con}\neq0
\label{e1}%
\end{equation}
where $T_{0}$ is defined in the same way as previous (take all derivatives in
front of the time-ordering). The violation of the $e~$and $e^{\prime}%
~$independence terms in the bracket are well-defined, and as in (\ref{an}) of
the third section the anomalies lead us to the necessary $N,R$ and $N_{\mu}$
modifications. The $L~$of the neutral model has more terms than massive scalar
QED, in fact it will turn out that the requirement of string-independence of
the S-matrix can only be fulfilled by adding yet another renormalizable
selfinteraction $cH^{4}.$ Again $b$ and$~c$ are induced couplings which turn
out to be proportional to $g^{2}$ and the masses $m,m_{H}$ of the two fields.
Despite the presence of more terms, the neutral scalar interactions depend,
just as its charged counterpart in the previous section, only on the massive
vectormeson-$H$ coupling $g.~$A new phenomenon is that the coefficients
$b,c~$also contain mass ratios of the masses of the two fields $m~$and
$m_{H}.$ In short: all terms beyond the basic $A\cdot AH$ interactions are
"induced" by the locality and positivity (Hilbert space) requirements of QFT.

It is helpful for the reader (and also matter of historical correction) to
give credit to previous important work which pointed to the misunderstandings
of the Higgs symmetry-breaking mechanism \cite{Scharf}\cite{Aste}. Their work,
which was based on the operator formulation of the BRST gauge setting,
unfortunately remained unnoticed.

In the BRST setting one starts from the counterpart of (\ref{e1}) in terms of
the abstract nilpotent $s$-operation replacing the concrete differential
calculus on the directional de Sitter space. The basic first order relation
which corresponds to (\ref{first}) is%
\begin{equation}
sL^{K}=\partial^{\mu}Q_{\mu}^{K}%
\end{equation}
where the $s$-operation on the field (including the ghost fields) was
mentioned at the end of the third section. In terms of this $s$ operation the
BRST anomaly is defined\footnote{Although we use the same notation $V_{\mu
},Q_{\mu}=sV_{\mu},$ they are very different operators in Krein space. Their
role with respect to the cohomological $s$ is analogies to the differential
cohomology in de Sitter space.}%
\begin{equation}
A^{K}:=(sT_{0}L^{K}L^{\prime K}-\partial^{\mu}T_{0}Q_{\mu}^{K}L^{\prime
K}-\partial^{\prime\mu}T_{0}L^{K}Q_{\mu}^{\prime K})_{1-con}%
\end{equation}
where $K$ refers to the Krein space and the violation of the $s$-invariance
comes again from delta function contributions which arise from the application
of the wave operator to time-ordered propagators. The $L^{K}$ and $Q_{\mu}%
^{K}~$in \cite{Scharf} are (for simplicity of notation we omit the superscript
$K$ for Krein space on the individual fields $A_{\mu},\phi,u$) of the BRST
gauge setting
\begin{align}
L^{K}  &  =m\left(  A\cdot AH-H\overleftrightarrow{\partial}\phi\cdot
A-\frac{m_{H}^{2}}{2m^{2}}H\phi^{2}+bH^{3}+u\tilde{u}H\right) \\
Q_{\mu}^{K}  &  =m(uA_{\mu}H-\frac{1}{m}u\phi\overleftrightarrow{\partial
}_{\mu}H)\nonumber
\end{align}
Apart from the appearance of the $u\tilde{u}~$ghost contribution and the
different engineering dimension of\ the negative metric St\"{u}ckelberg field
$\phi^{K}$\ as compared to the stringlocal escort $\phi~$($m\phi\sim\phi^{K}$)
the formulas are identical, despite the big difference between the conceptual
aspects of their derivation.$.$

There are now two propagators
\begin{align}
\partial^{\mu}\left\langle T_{0}\partial_{\mu}H\partial_{\nu}^{\prime
}H^{\prime}\right\rangle  &  =-i\partial_{\nu}^{\prime}\delta(x-x^{\prime
})+reg,~~\partial^{\mu}\left\langle T_{0}\partial_{\mu}HH^{\prime
}\right\rangle =-i\delta(x-x^{\prime})+reg~\\
\partial^{\mu}\left\langle T_{0}\partial_{\mu}\phi\partial_{\nu}^{\prime}%
\phi^{\prime}\right\rangle  &  =i\partial_{\nu}^{\prime}\delta(x-x^{\prime
})+reg,~~\partial^{\mu}\left\langle T_{0}\partial_{\mu}\phi\phi^{\prime
}\right\rangle =i\delta(x-x)+reg
\end{align}
The negative sign in the second line comes from the Krein field $\phi$ whose
two-point function has the opposite sign (the negative metric St\"{u}ckelberg
field). Again we use the freedom of normalization
\begin{align}
\left\langle T\partial_{\mu}H\partial_{\nu}^{\prime}H^{\prime}\right\rangle
&  =\left\langle T_{0}\partial_{\mu}H\partial_{\nu}^{\prime}H^{\prime
}\right\rangle -ia_{H}g_{\mu\nu}\delta(x-x^{\prime})\\
\left\langle T\partial_{\mu}\phi^{K}\partial_{\nu}^{\prime}\phi^{\prime
K}\right\rangle  &  =\left\langle T_{0}\partial_{\mu}\phi^{K}\partial_{\nu
}^{\prime}\phi^{\prime K}\right\rangle +ia_{\phi}g_{\mu\nu}\delta(x-x^{\prime
})\nonumber
\end{align}
As in the previous section, the contribution to the $N$ of the anomaly coming
from contractions of $\partial_{\mu}H~~$and $\partial_{\mu}\phi~$with the
first two terms in $L$ can be absorbed in a redefinition $T_{0}\rightarrow
T~$made to vanish by choosing $a_{H}=1=a_{\phi}.~$These terms lead to a
nontrivial $R$-contribution to the anomaly (\ref{an})$~$

But in contrast to the massive QED case, the story does not end here. There
are two $A_{\mu}$-independent quadrilinear remaining delta anomaly terms which
result from contractions of the second term in $Q_{\mu}^{K}$ with the third
and fourth term in $L^{K}.$ They lead to a "potential" in the two scalar
fields
\begin{equation}
R=-i\delta(x-x^{\prime})\left\{  -(\frac{m_{H}^{2}}{m^{2}}+3b)\phi^{2}%
H^{2}+\frac{m_{H}^{2}}{4m^{2}}\phi^{4}\right\}
\end{equation}
As in the case of massive QED we omit the calculation of the $N_{\mu}$ which
renormalize the time order products $Q_{\mu}L^{\prime}$ and $LQ_{\mu}^{\prime
}.$ At this point the $b$ is still a free coupling parameter.

In order to get from the $R$-potential to the Mexican hat form, we follow
Scharf \cite{Scharf} and observe that the tree approximation of the third
order has a nontrivial anomaly which comes from the time ordered product the
first order $Q_{\mu}^{k}~$with the second order potential $T_{0}Q_{\mu}%
R.~$Without adding the before mentioned $cH^{4}$\ to the induced potential
$R~$it is not possible to get rid of this anomaly. With this term one finds
compensation for the following values of $b,c$%
\begin{align}
b  &  =-\frac{m_{H}^{2}}{2m^{2}},~c=-\frac{m_{H}^{2}}{4m^{2}}\\
R_{ind}^{K}  &  =-i\delta(x-x^{\prime})\frac{m_{H}^{2}}{4m^{2}}(H^{2}+\phi
^{2})^{2}%
\end{align}

As in the case of massive QED one may combine the induced second order
$R_{ind}~$potential with the $\phi,H~$dependent part of the first order and
write the result in the form
\begin{align*}
V_{1}^{K}  &  =g\frac{m_{H}^{2}}{2m}(H\phi^{2}+H^{3}),~V_{2}=g^{2}\frac
{m_{H}^{2}}{4m^{2}}(H^{2}+\phi^{2})^{2}\\
V^{K}  &  =V_{1}^{K}+\frac{1}{2}V_{2}^{K}=\frac{m_{H}^{2}}{8m^{2}}(H^{2}%
+\phi^{2}+\frac{2m}{g}H)^{2}-\frac{m_{H}^{2}}{2}H^{2}\\
S^{K}  &  =1+i\int gA\cdot AH-\int\int(\delta(x-x^{\prime})V+\frac{g^{2}}%
{2}TLL^{\prime})+higher~order
\end{align*}
An obvious c-number field shift in the $H$-field leads to the symmetric form
of the Mexican hat potential (the $V$ without the mass term). But there is no
physical reason for writing the induced potential in this form.

The important message here is that the requirement of second order gauge
independence of $S$ in the form of $sS^{(2)}=0~$uniquely$~$determines the
changes of $T_{0}LL^{\prime}~$which must be done in order to achieve this
task. There is no symmetry-breaking or mass generation, rather the model is
defined in terms of the original trilinear interaction between the massive
vectorpotential and a Hermitian field. The Mexican hat potential is induced by
this elementary interaction; it is not part of the definition of the model as
in the case of the unphysical Higgs mechanism.

Second order calculations contain no information about quadrilinear
self-couplings which are not induced (counterterms whose couplings define new
parameters). One would expect that the 4$^{th}~$order box-contributions lead
to additional self-couplings \thinspace$(\varphi^{\ast}\varphi)^{2}$ in QED
and $H^{4}~$in the H-model and, unlike induced potentials, are genuine
undetermined renormalization parameters. This problem will be taken up in a
separate publication \cite{M-S1}.

The correct formulation of the Higgs model in terms of a $A$-$H~$coupling
shows that there is a irreconcilable difference with a Goldstone spontaneous
symmetry breaking. Whereas the conserved current in the Goldstone situation
leads to a diverging charge $Q_{G}=\infty$ (this is the intrinsic definition
of spontaneous symmetry breaking), the only conserved current in the Higgs
model is screened $Q_{H}=0$. In fact its screening is a special case of a
property shared by all theories which couple massive vectormesons to matter
(the Schwinger-Swieca screening). Between these two extremes there is the
normal symmetry situation in which the $Q$ is finite and nontrivial.

As mentioned before the critique of the Higgs mechanism points to a radical
change in the conceptual relation between massive and massless interacting
fields. A much stronger indication that massless $s\geq1$ interactions should
be constructed as limits of massive models with their clear field-particle
relation comes from the still unsolved problems of QCD confinement and the
only partially understood problem of QED infraparticles. Here the stringlocal
Hilbert space setting is essential since the unphysical pointlike objects of
the the BRST gauge setting are too removed from the physical matter fields
which are necessarily stringlocal.

In the following we will assume that we are in the zero mass situation where
singular pointlike descriptions of matter do not exist. In this case it is
reasonable to distinguish between reducible and irreducible stringlocal
fields. The former are fields which can be described as long distance limits
of finite localized objects, whereas for irreducible fields this is not
possible. Among free fields only the noncompact third Wigner positive energy
class is irreducible: no stringlocal field associated to such a representation
(not even composites \cite{MSY}) can be represented as integrals over
pointlike fields. Interacting abelian potentials are integrals over observable
field strengths, but it is not possible to represent interacting Y-M gluon
fields in this way. For such inherently noncompact objects their creation from
collisions of compact matter leads to problems with the principle of causal
localization. Whereas causality forces third class Wigner matter, in case it
occurs in our universe, to be maximally "inert" cosmological staff without any
possibility to interact apart from gravity (dark matter ?, see section 2), it
prevents interacting gluons from emerging from a collision of ordinary matter
in a compact region.

Formulated in terms of massless limits of correlations of massive stringlike
gluons this strongly suggests to define gluon confinement as the
\textit{vanishing of the limiting correlations functions} which contain in
addition to pointlike composites also gluon or quark operators; in this way
the mentioned causality problems disappear. For quarks in QCD which carry a
charge, one expects the only exception for configurations in which the string
direction $e~$of $q$-$\bar{q}$\ are parallel to the spacelike distance of the
endpoints. The benefit of this definition of confinement is that it suggests a
way to prove it by resummation of the leading logs in the limiting correlation
functions in which the $m$ is used as a natural infrared regularization
parameter. The historically educated reader will recall that a similar
resummation idea was used by Yenni-Frautschi-Suura in order to show that the
scattering amplitude for charged particles with only a finite number of photon
vanish (the only nontrivial scattering data a photon-inclusive cross
sections). These calculations were done in terms of resummation of leading
logarithmic infrared contributions of infrared regularized scattering
amplitudes; the present setting suggests to re-do these calculations with the
use of the limiting vectormeson mass as natural covariant infrared regulator.
The before defined gluon/quark confinement should result from similar
resummation of leading infrared terms in \textit{off-shell} massive
gluon/quark correlations for $m\rightarrow0.$ which plays a similar role as
the ad hoc regularization parameter in the argument which implies the
vanishing of charged particle scattering with a \textit{finite} number of
outgoing photons. The difference is that the QED infrared problem is on-shell
(the electron string is reducible ), whereas confinement is an off-shell phenomenon.

The cited work in \cite{Scharf}\cite{Aste} in which the Higgs mechanism was
replaced by a $A$-$H~$coupling, as well as the present Hilbert space
formulation of higher spin $s\geq1~$field interactions are rather late
attempts to point at problems which, despite their more than 40 years of
existence, need more conceptual attention. Recalling that Swieca's effort to
direct the focus of attention away from symmetry-breaking by using the
terminology "Schwinger-Higgs" in most of his publications \cite{Swieca}
("Schwinger" for the screening idea and "Higgs" for the neutral $H$-coupling)
got lost in the maelstrom of time, one cannot be optimistic about the success
of the present attempts to shed additional critical light on these old
problems. This is particularly difficult if incomplete results, which are
still not anywhere near to their closure, have been sanctioned by Nobel prizes.

\section{Resum\'{e} and outlook}

New concepts, which shed light on insufficiently understood old problems,
usually lead to new questions, and the extension of QFT to string-localized
fields is no exception. The clarification of the old controversies about
spontaneous symmetry breaking and mass generation in this paper was obtained
with a rather modest computational effort within a new setting. The new
concepts used to achieve this show that Hilbert space positivity requires a
quite different formalism from that of pointlike fields. Whereas in the latter
case the perturbative systematics can be encoded into Feynman rules for which
the different type of vertices represent independent couplings, such graphical
presentations loose their utility in the presence of \textit{induced}
normalization contributions with computable coupling strengths. Our new
Hilbert space setting sheds very different light on open problems related to
the Higgs model. Its systematic mathematical presentation requires a
nontrivial extension of the locality based inductive Epstein-Glaser
renormalization formalism \cite{E-G} from point to string crossings
\cite{Jens1}.

The fact that there exist stringlocal with the minimal short distance
dimension for all spins\footnote{d=1 for integer spin and d=3/2 for fermionic
strings.}(d=1 for integer spin and d=3/2 for fermionic strings) permits to
define interactions which remain within the power-counting criterion of
renormalization theory for all spins. The appearance of a $s=0~$lower spin
Higgs-like intrinsic escort field $\phi$ for $s=1~$is a special case of a new
phenomenon, namely the presence of $s$ stringlocal intrinsic escort fields of
lower spin (in the fermionic case there are $s-1/2$ lower spin spinor fields)
which are inexorably linked to the massive stringlocal spin $s$ field and
appear explicitly in its interaction. The presence of pointlocal observables
is an additional physical restriction on interactions. It is truly surprising
that for the rather small prize of weakening locality from point- to
stringlike one is able to re-opens a whole new world of $s\geq1$
renormalizable models of QFT.

There are important problems for $s=1$ which cannot be properly addressed in
the BRST gauge setting. The matter fields of the gauge setting are unphysical,
the only renormalizable physical matter fields\footnote{As explained in this
paper, one can define "singular pointlike" fields in terms of the
renormalizable stringlike fields which are well-defined in every order, but
they are not renormalizable in the standard sense.} are stringlocal and hence
outside the range of gauge theory (this also includes physical selfinteracting
Y-M fields). The new Hilbert space setting as presented in this paper
addresses problems of massive $s\geq1~$fields; massless situations have to be
approached by taking massless limits of massive correlation functions; the
latter can then be used to reconstruct a zero mass operator QFT \cite{St-Wi}.
There are a good physical reasons for approaching massless situations from the
massive side; the physics behind massless interaction remains largely unknown.
It hides phenomena as gluon/quark confinement as well as incompletely
understood infrared aspects of charged infraparticles. The new setting creates
favorable conditions for their solution in that the stringlike physical fields
incorporate the very restrictive Hilbert space positivity.

Looking back at history and recalling that the idea of the Higgs mechanism
originated from a time in which massless models, as QED, were considered to be
simpler than their massive counterpart, the new message supports the opposite
view. Whereas renormalizable couplings of massive vectormesons to charged or
neutral matter (as well as Y-M self-couplings) lead to standard field-particle
picture backed up by scattering theory, all this breaks down for $s\geq1$ in
the massless limit. Massless models present the real challenge, and there is
the good chance to understand them in the new SLF Hilbert space setting in
terms of massless limits of physical objects.

Taking a more philosophical stance, one may say the new setting de-mystifies
the gauge principle in favor of substituting it by the foundational causal
\textit{localization principle in Hilbert space; in this way all models of
QFT, independent of the spin of their fields, are unified under the shared
conceptual roof of the causal localization principle.}

\textbf{Acknowledgement}: I am indebted to Jens Mund for numerous discussions
and for making his notes available prior to publication and last not least for
reading my manuscript and suggesting improvements. I also thank Raymond Stora
for having accompanied this new development with great interest and
challenging questions which led to better formulations of some important points.

\end{document}